\documentclass{aa}

\usepackage{rotating}
\usepackage{natbib}
\usepackage{mathtools}
\usepackage{amsmath}
\usepackage[normalem]{ulem}
\usepackage[percent]{overpic}

\usepackage[dvipsnames]{xcolor}

\DeclarePairedDelimiter\abs{\lvert}{\rvert}%
\makeatletter
\let\oldabs\abs
\def\abs{\@ifstar{\oldabs}{\oldabs*}}

\begin{document}

\title{Mode conversion of two-fluid shocks in a partially-ionised, isothermal, stratified atmosphere.}
\titlerunning{PIP resonance}

\author{B. Snow\inst{1}, A. Hillier\inst{1}}
\institute{University of Exeter, Exeter, EX4 4QF, UK \email{b.snow@exeter.ac.uk} \label{inst1}}


\abstract 
{The plasma of the lower solar atmosphere consists of mostly neutral particles, whereas the upper solar atmosphere is mostly ionised particles and electrons. A shock that propagates upwards in the solar atmosphere therefore undergoes a transition where the dominant fluid is either neutral or ionised. An upwards propagating shock also passes a point where the sound and Alfv\'en speed are equal. At this point the energy of the acoustic shock can separated into fast and slow components. How the energy is distributed between the two modes depends on the angle of magnetic field.}
{The separation of neutral and ionised species in a gravitationally stratified atmosphere is investigated. The role of two-fluid effects on the structure of the shocks post-mode-conversion and the frictional heating is quantified for different levels of collisional coupling.}
{Two-fluid numerical simulations are performed using the (P\underline{I}P) code of a wave steepening into a shock in an isothermal, partially-ionised atmosphere. The collisional coefficient is varied to investigate the regimes where the plasma and neutral species are weakly, strongly and finitely coupled.}
{The propagation speeds of the compressional waves hosted by neutral and ionised species vary, therefore velocity drift between the two species is produced as the plasma attempts to propagate faster than the neutrals. This is most extreme for a fast-mode shock. We find that the collisional coefficient drastically changes the features present in the system, specifically the mode conversion height, type of shocks present, and the finite shock widths created by the two-fluid effects. In the finitely-coupled regime fast-mode shock widths can exceed the pressure scale height leading to a new potential observable of two-fluid effects in the lower solar atmosphere.
}
{}


\maketitle

\section{Introduction}

A compressional wave propagating upwards in the solar atmosphere naturally steepens due to the stratification of the atmosphere and can readily develop nonlinearities and shock \citep[e.g.,][]{Suematsu1982}. This process occurs throughout the solar atmosphere, for example, it is thought to be the cause of umbral flashes \citep{Beckers1969,Houston2018,Tetsu2019}. These waves are magnetohydrodynamic (MHD) in nature and thus depend on the inclination of the magnetic field. 

Mangetohydrodynamic (MHD) systems are capable of supporting numerous types of shock transitions due to the three characteristic speeds (slow, Alfv\'en, fast) \cite[e.g.,][]{Delmont2011}. Slow-mode shocks feature a transition from superslow to subslow and are of particular importance for magnetic reconnection events \citep{Petschek1964,Innocenti2015,Shibayama2015}. Fast-mode shocks propagate perpendicular to the magnetic field and are also important for magnetic reconnection \citep{Yokoyama1996}. There in also a class of intermediate shocks which have a transition from above to below the Alfv\'en speed, which can arise due to two-fluid effects in slow-mode shocks \citep{Snow2019}.

It is well known that MHD waves can undergo mode conversion in the solar atmosphere \citep[e.g.,][]{Schunker2006,Cally2011,Khomenko2019}, and the same is true of shocks. A recent paper by \cite{Pennicott2019} studied the propagation of an MHD shock in a stratified, isothermal atmosphere and the separation into slow- and fast-mode shocks above the $\beta _v$ point, which is defined as the height in the atmosphere where the sound and Alfv\'en speeds are equal. Their results showed that as a shock passes through the $\beta _v$ point it can can undergo mode conversion. In cases where the angle between the magnetic field and the vertical is large, the slow component of the shock formed after mode conversion becomes smoothed.

Many events involving shocks can occur in the lower solar atmosphere where both neutral and ionised species are present, and the interactions between these species are thought to play a key role in the underlying dynamics.
The lower solar atmosphere is partially ionised, being mostly neutral fluid in the photosphere and chromosphere, and mostly ionised above the transition region. 
Partial ionisation is known to have several physical consequences in this region including additional heating through dissipation \citep[e.g.][]{Khomenko2012b} and enhanced reconnection rates \citep{Leake2013}. Studies of two-fluid waves in a stratified atmosphere show that the two-fluid effects lead to additional damping \citep{Popescu2019}. 
Reviews of partial ionisation can be found in \cite{Khomenko2017,Ballester2018,Khomenko2020}.

Shocks in partially-ionised plasma (PIP) are of particular interest due to the decoupling and recoupling of species around the shock front \citep{Hillier2016}. As such, shocks are a promising area for the two-fluid effects to have a significant impact on the physics of the system. Previous work has shown that additional shocks can form within the finite width of a reconnection-driven slow-mode shock, \citep{Snow2019}.

In this paper, we investigate the effect of gravitational stratification and partial ionisation on a propagating shock initiallised through wave steepening. In particular, we investigate the role of two-fluid effects on mode conversion of the shock across the $\beta _v$ point and the consequences of two-fluid interactions on the structure of the propagating shocks. The collisional coefficient is critical in determining the mode conversion point, the type of shocks present, and the width of these shocks. The fast-mode shock width can exceed the pressure scale height for a broad range of coupling coefficients, which we present as a new potential observable of two-fluid effects in the lower solar atmosphere. 

\section{Numerical methods}

The numerical simulations are performed using the (P\underline{I}P) code \citep{Hillier2016} to solve the evolution of a two fluid (neutral and ion+electron) hydrogen plasma. 
The non-dimensional equations for the evolution of the neutral (subscript $n$) and charge-neutral plasma (subscript $p$) used in our study are given by:
\begin{gather}
\frac{\partial \rho _{\text{n}}}{\partial t} + \nabla \cdot (\rho _{\text{n}} \textbf{v}_{\text{n}})=0, \label{eqn:neutral1} \\
\frac{\partial}{\partial t}(\rho _{\text{n}} \textbf{v}_{\text{n}}) + \nabla \cdot (\rho _{\text{n}} \textbf{v}_{\text{n}} \textbf{v}_{\text{n}} + P_{\text{n}} \textbf{I}) \nonumber \\  \hspace{0.5cm} = \rho_{\text{n}} \textbf{g} -\alpha _c \rho_{\text{n}} \rho_{\text{p}} (\textbf{v}_{\text{n}}-\textbf{v}_{\text{p}}), \\
\frac{\partial e_{\text{n}}}{\partial t} + \nabla \cdot \left[\textbf{v}_{\text{n}} (e_{\text{n}} +P_{\text{n}}) \right] \nonumber \\  \hspace{0.5cm}= \rho_{\text{n}} \textbf{g} \cdot \textbf{v}_{\text{n}} -\frac{\alpha _c \rho _{\text{n}} \rho _{\text{p}}}{2} \left[ (\textbf{v}_{\text{n}} ^2 - \textbf{v}_{\text{p}} ^2)+ 3 \left(\frac{P_{\text{n}}}{\rho_{\text{n}}}-\frac{1}{2}\frac{P_{\text{p}}}{\rho_{\text{p}}}\right) \right], \label{eqn:energyn} \\
e_{\text{n}} = \frac{P_{\text{n}}}{\gamma -1} + \frac{1}{2} \rho _{\text{n}} v_{\text{n}} ^2, \label{eqn:neutral2}\\
\frac{\partial \rho _{\text{p}}}{\partial t} + \nabla \cdot (\rho_{\text{p}} \textbf{v}_{\text{p}}) = 0 \label{eqn:plasma1}\\
\frac{\partial}{\partial t} (\rho_{\text{p}} \textbf{v}_{\text{p}})+ \nabla \cdot \left( \rho_{\text{p}} \textbf{v}_{\text{p}} \textbf{v}_{\text{p}} + P_{\text{p}} \textbf{I} - \mathbf{B B} + \frac{\textbf{B}^2}{2} \textbf{I} \right) \nonumber \\  \hspace{0.5cm}= \rho _{\text{p}} \textbf{g} + \alpha _c \rho_{\text{n}} \rho_{\text{p}}(\textbf{v}_{\text{n}} - \textbf{v}_{\text{p}}), \\
\frac{\partial}{\partial t} \left( e_{\text{p}} + \frac{\textbf{B}^2}{2} \right) + \nabla \cdot \left[ \textbf{v}_{\text{p}} ( e_{\text{p}} + P_{\text{p}}) -  (\textbf{v}_p \times \textbf{B}) \times \textbf{B} \right] \nonumber \\  \hspace{0.5cm} = \rho_{\text{p}} \textbf{g} \cdot \textbf{v}_{\text{p}} + \frac{\alpha _c \rho _{\text{n}} \rho _{\text{p}}}{2} \left[  (\textbf{v}_{\text{n}} ^2 - \textbf{v}_{\text{p}} ^2)+ 3 \left(\frac{P_{\text{n}}}{\rho_{\text{n}}}-\frac{1}{2}\frac{P_{\text{p}}}{\rho_{\text{p}}}\right) \right], \label{eqn:energyp}\\
\frac{\partial \textbf{B}}{\partial t} - \nabla \times (\textbf{v}_{\text{p}} \times \textbf{B}) = 0, \\
e_{\text{p}} = \frac{P_{\text{p}}}{\gamma -1} + \frac{1}{2} \rho _{\text{p}} v_{\text{p}} ^2, \\
\nabla \cdot \textbf{B} = 0,\label{eqn:plasma2}
\end{gather}
for density $\rho$, velocity $\textbf{v}$, pressure $P$, internal energy $e$, magnetic field $\textbf{B}$, the collisional coupling term $\alpha_c$, and gravity $\textbf{g}$. Note that the thermal component of the collisional term in Equations (\ref{eqn:energyn}) and (\ref{eqn:energyp}) is different from in previous papers using the (P\underline{I}P) code by a factor of $1/2$ \citep{Hillier2016,Snow2019}. The corrected term was found to have no significant changes for the reconnection driven shock considered in previous works. The thermal coupling term is now consistent with other two-fluid numerical codes, e.g., \cite{Popescu2019}.

This system of equations has been non-dimensionalised in the following way: The velocity $\mathbf{v}$ has been non-dimensionalised using the sound speed of the plasma fluid $c_{\rm s}$, the density $\rho$ by a reference plasma density $\rho_{\rm{p,0}}$, and the lengthscale by the characteristic pressure scale height of the plasma $\Lambda_{\rm{p,0}}$.
This implies that the timescales of the system are nondimensionalised by $\Lambda_{\rm{p,0}}c_{\rm s}^{-1}$, the pressure $P$ by $c_{\rm s}^2\rho_{\rm{p,0}}$ and the magnetic field $\mathbf{B}$ by $B_0/\sqrt{4\pi}=c_{\rm s}\sqrt{\rho_{\rm{p,0}}}$. 
Here the subscript $0$ is used to represent a reference value of a quantity (where these values have been chosen to be those from the $x=0$ point) and $\gamma$ is the adiabatic index.
Non-dimenstionalisation using the plasma components was selected as it means that the results of reference MHD simulations would be exactly reproduced in the case of $\alpha_c=0$.
We assume that both fluids follow the ideal gas law which for the two fluids in non-dimensional form become $T_{\rm n}=\tfrac{P_{\rm n}\gamma}{\rho_{\rm n}}$ and $T_{\rm p}=\tfrac{P_{\rm p}\gamma}{2\rho_{\rm p}}$, respectively.

Here we use the first order HLLD solver in the (P\underline{I}P) code to accurately capture shocks. 
The HLLD scheme is described in detail in \cite{Miyoshi2005} and can accurately model shocks. This scheme has been used previously with the (P\underline{I}P) code to model MHD and two-fluid shocks across a range of plasma-$\beta$ values and ionisation fractions \citep{Hillier2016,Snow2019}. A discussion of the numerical diffusion of the scheme is included in Appendix \ref{sec:numdif}.
The domain is spatially resolved by 8000 grid cells.
The equations are solved for 1.5D evolution in that there is variation in the $x$ direction, but velocities and magnetic fields can exist in the $y$ direction.

\subsection{Initial conditions}

\begin{figure}
    \centering
    \includegraphics[width=0.5\textwidth,clip=true,trim=0.8cm 7.5cm 1cm 7cm]{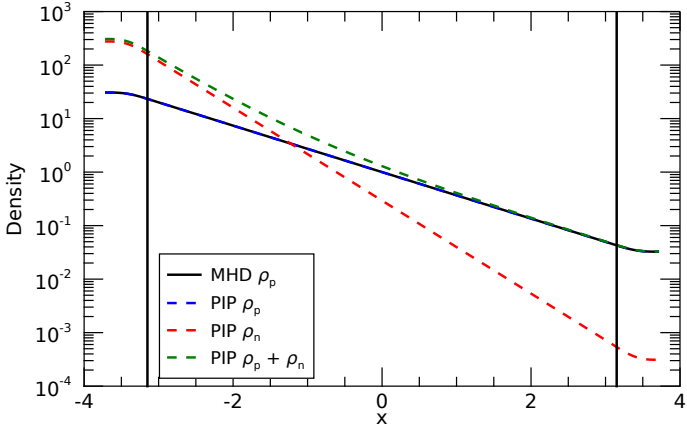}
    \caption{Density for the initial equilibrium. MHD density is the black-solid line. The PIP simulation has two densities: plasma (blue dashed) and neutral (red dashed). The normalisation leads to the MHD density being the same as the PIP plasma density. The total density for the PIP case is shown in green-dashed. Gravity is constant between the two vertical black lines.}
    \label{fig:densities}
\end{figure}

In this paper, we consider an isothermal atmosphere to study the propagation of two-fluid shocks in an simplified system. The initial density profile for the MHD and PIP simulations is shown in Figure \ref{fig:densities}.
This initial set up is similar to that of \cite{Pennicott2019} except here we have the addition of two-fluid effects. Future work will include studying shock behaviour in more realistic atmospheres.

At the simulation boundaries, gravity rolls down to zero. This is purely to stabilise the boundary and simplify the driver formulation. 
This has no consequence on the propagation of the waves. Gravity is applied in the $x$ direction only ($\textbf{g}=[g_x,0,0]$), where $g_x$ is specified as a piecewise function of height as:
\[ 
g_x= \left\{
\begin{array}{ll}
      0 & x \leq g_1 \\
      g_0 \left(1+{\rm cos}\left(\pi \frac{x-g_2}{g_1-g_2}\right)\right) & g_1 \leq x\leq g_2\\
      g_0 & g_2 \leq x \leq g_3\\
      g_0 \left(1+{\rm cos}\left(\pi \frac{x-g_3}{g_4-g_3}\right)\right) & g_3\leq x\leq g_4 \\
      0 & g_4 \leq x \\
\end{array} 
\right. 
\]
where $g_0 = 1/\gamma$ from the normalisation. $g_1=-3.7,g_4=3.7$ are the upper and lower grid limits, such that gravity is zero in our ghost cells. $g_2 =-3.15$ and $g_3=3.15$ and between these limits the gravity is constant. The region where gravity is constant is marked on Figure \ref{fig:densities} by the vertical black lines.

Our equilibrium pressure and density profiles are calculated by integrating upwards using the pressure scale heights for the two species, with the density calculated from the ideal gas:
\begin{gather}
    P_{\rm{p;n}}(x) = P_{\rm{p;n}}(0) \exp \left(-\int_{g_1}^{x} \frac{1}{\Lambda _{\rm p;n}}  dx'\right), \\
    \Lambda _{\rm{n}} = \frac{T_{\rm{n}}}{\gamma g_x}, \\
    \Lambda _{\rm{p}} = \frac{2T_{\rm{p}}}{\gamma g_x}.
\end{gather}
Normalisation is chosen such that the plasma density is unity at $x=0$. Initially $T_{\rm{p}}=T_{\rm{n}}=1/2$ . All simulations have the same plasma density profile, which results in the PIP simulations having greater total mass, see Figure \ref{fig:densities}. 
We set the magnetic field strength as $|\textbf{B}| = 1$ such that the $\beta_v$ point is at $x=0$.

The non-dimensional collisional terms are calculated using the average temperature:
\begin{gather}\label{drift}
    \alpha _c = \alpha _0 \sqrt{\frac{T_{\rm n}+T_{\rm p}}{2}}.
\end{gather}
The collisional coupling coefficient is set to $\alpha _0 = 100$ unless otherwise stated.

\subsection{Boundary conditions}

A pulse of one half period is applied to the system at the lower boundary of the form:
\begin{gather}
    v_{x,{\rm p;n}}  = A \sin{(2 \pi t / \tau)}, \label{eqn:driver1} \\
    \rho_{\rm p;n} = A \frac{\rho_{{\rm p;n}}(-3.7)}{c_s} \sin{(2 \pi t / \tau)}, \\ 
    P_{\rm p;n}    = A \frac{\gamma P_{{\rm p;n}}(-3.7)}{c_s} \sin{(2 \pi t / \tau)},    \label{eqn:driver3}
\end{gather}
for period $\tau$, initial density $\rho_{-3.7}$, initial pressure $P_{-3.7}$ at the base of the domain, and initial sound speed $c_s$. This form of driver has the advantage of preventing additional modes being driven, compared to exciting velocity only, and is equivalent to the drivers used in \citet{Khomenko2012} and \citet{Felipe2019}. For details on the numerical implementation see Appendix \ref{driver}.

The period is chosen as $\tau = 0.5$ in normalised time units, and the normalised wave amplitude $A=0.1$. The choice of amplitude is such that the wave steepens into a shock below the $\beta _v$ point (where $c_s=v_A$). The acoustic cutoff frequency can be calculated using $\omega _c = \frac{c_s}{2H} \sqrt{1-2\frac{dH}{dx}}$, where $H=\frac{c_s^2}{\gamma g}$. Where gravity is constant ($-3.15<x<3.15$), $\omega _c = 0.5$ which is smaller than the driver frequency of $2\pi/ \tau = 4 \pi$. Due to the gravity variations near our boundaries the acoustic cutoff varies near the edges of our domain however the maximum value is $\max (\omega _c) \approx 1.22 $ which is still smaller than the driver frequency.

In the two-fluid case, the velocity is applied to both the neutral and plasma species. As such, more total kinetic energy is applied to the two-fluid case than the MHD case since the PIP simulation have higher total density, see Figure \ref{fig:densities}. 

The boundary at the upper end of the domain is taken to be an open boundary. Simulations are stopped before the fast-mode component reaches this boundary.

\section{Fundamentals of two-fluid waves}

\subsection{Characteristic speeds in two-fluid plasma} \label{sec:wavespeeds}

For two-fluid plasma, one can define the wave speeds based on the total density, or the plasma density alone:
\begin{gather}
    c_s^2={\gamma P_{\rm p}/\rho_{\rm p}}, \\
    c_{sn}^2={\gamma P_{\rm n}/\rho_{\rm n}}, \\
    c_{st}^2={\gamma (P_{\rm n}+P_{\rm p})/(\rho_{\rm n} +\rho_{\rm p})}, \\
    v_A^2={(B_x^2+B_y^2)/\rho}, \\
    v_{At}^2={(B_x^2+B_y^2)/(\rho _{\rm p} +\rho _{\rm n})}, \\
    \hat{v}_{A} = v_{At} \cos (\theta), \\
    \theta _B =\arctan (B_y/B_x), \\
    V_s^2={\frac{1}{2}\left(c_s^2+v_A^2 - \sqrt{(v_A^2+c_s^2)^2 - 4 v_A^2 c_s^2 \cos(\theta_B)^2)}\right)}, \\
    V_{st}^2={\frac{1}{2}\left(c_{st}^2+v_{At}^2 - \sqrt{(v_{At}^2+c_{st}^2)^2 - 4 v_{At}^2 c_{st}^2 \cos(\theta_B)^2)}\right)}, \\
    V_f^2={\frac{1}{2}\left(c_s^2+v_A^2 + \sqrt{(v_A^2+c_s^2)^2 - 4 v_A^2 c_s^2 \cos(\theta_B)^2)}\right)}, \\
    V_{ft}^2={\frac{1}{2}\left(c_{st}^2+v_{At}^2 +\sqrt{(v_{At}^2+c_{st}^2)^2 - 4 v_{At}^2 c_{st}^2 \cos(\theta_B)^2)}\right)},
\end{gather}
for the sound $c_s$, Alfv\'en $v_A$, slow $V_s$ and fast $V_f$ wave speeds. $\theta _B$ is the angle of the magnetic field. $\hat{v}_{A}$ is the component of the Alfv\'en velocity parallel to the $x$-axis and is necessary to calculate the shock jump conditions for intermediate shocks (see Appendix \ref{transistions} for details).

For an entirely coupled system ($\alpha _c \rightarrow \infty$), the wave speeds depend on the total density ($\rho _n + \rho _p$). For a entirely decoupled system ($\alpha _c = 0$) the wave speeds in the plasma depend on only the plasma density ($\rho _{\rm p}$) and those in the neutrals depend on only the neutral density ($\rho _{\rm n}$). 
Here we are interested in the region where a propagating wave undergoes a finite number of collisions as it crosses a pressure scale height. In this situation the majority of the velocity drifts are expected to exist in the shock front.

A dispersion relation exists that can be numerically solved to calculate the effective wave speeds for a finitely coupled plasma \cite{Soler2013}. We do not take this approach in this paper and instead consider the wave speeds at the extremes of coupling defined by the above equations.

\begin{figure}
    \centering
    \includegraphics[width=0.5\textwidth,clip=true,trim=1.8cm 8.5cm 2.11cm 9.1cm]{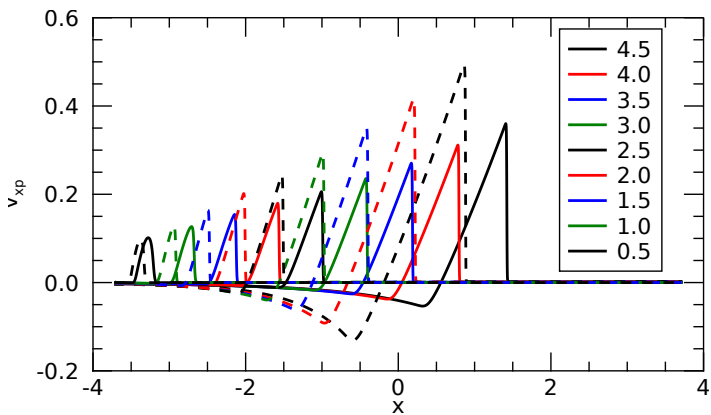}
    \caption{Snapshots showing the wave steepening for the MHD (solid) and the plasma component of the PIP (dashed) cases with a vertical magnetic field at different times. }
    \label{fig:wavesteepen}
\end{figure}

\begin{figure}
    \centering
    \includegraphics[width=0.5\textwidth,clip=true,trim=2.6cm 8.5cm 2cm 9cm]{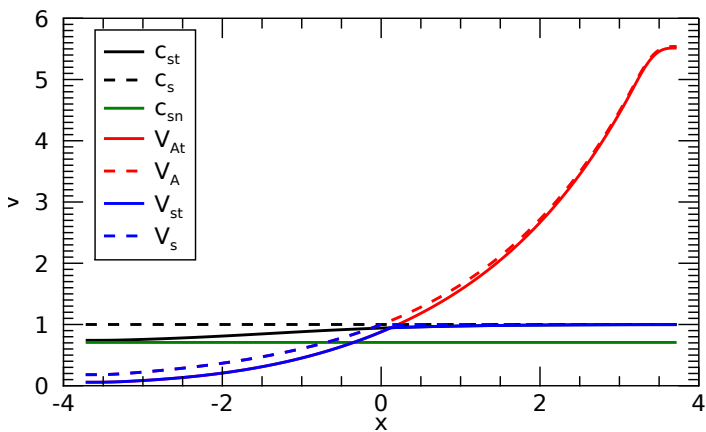}
    \caption{Wave speeds for a PIP simulation with a vertical magnetic field. The sound (black), Alfv\'en (red) and slow (blue) speeds are shown for the initial atmosphere using the total densities (solid) and plasma density only (dashed).} 
    \label{fig:wavespeeds}
\end{figure}

\begin{figure*} 
    \centering
    \includegraphics[width=0.49\textwidth,clip=true,trim=5.85cm 8cm 5.15cm 3.2cm]{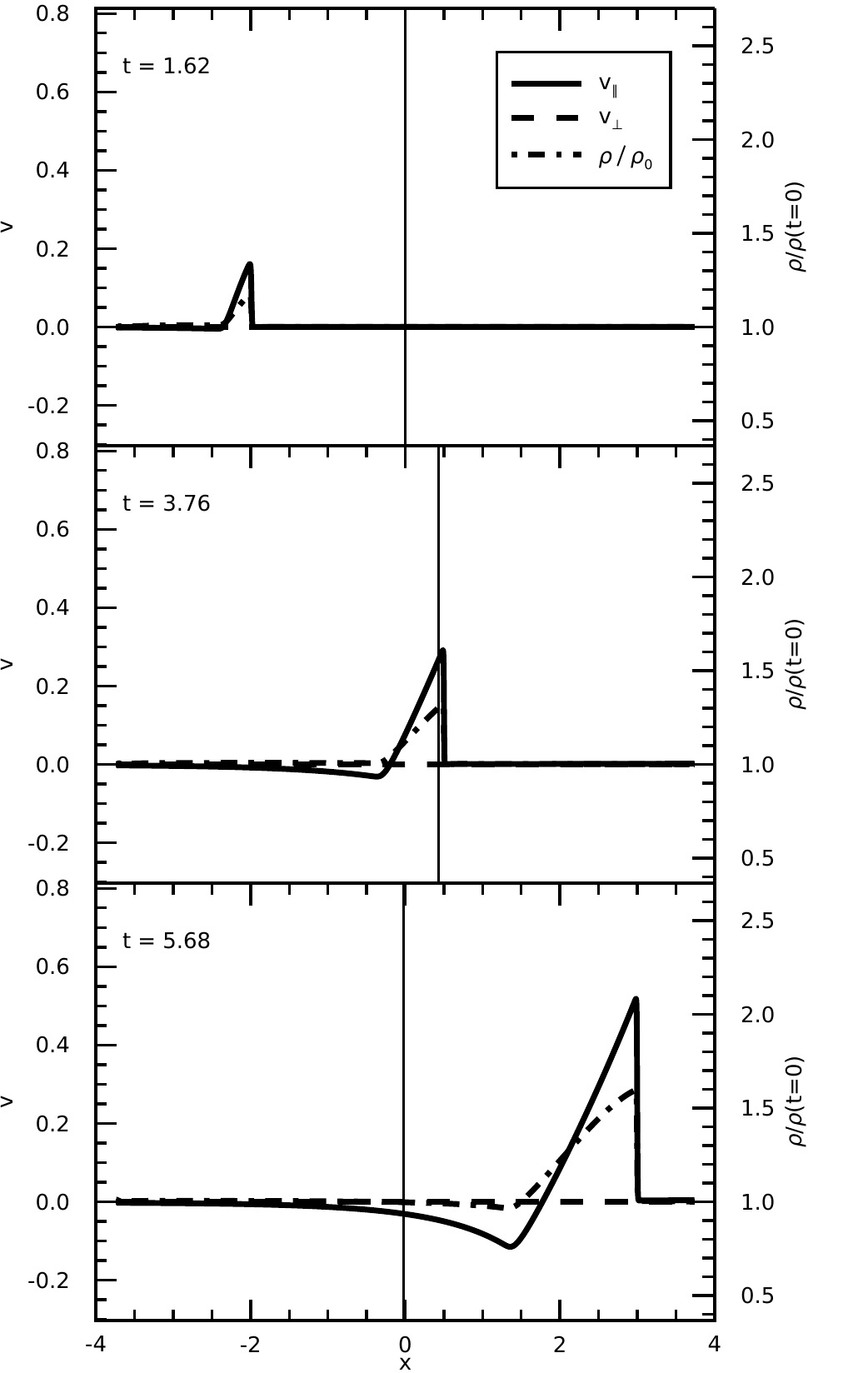} 
    \includegraphics[width=0.49\textwidth,clip=true,trim=5.85cm 8cm 5.15cm 3.2cm]{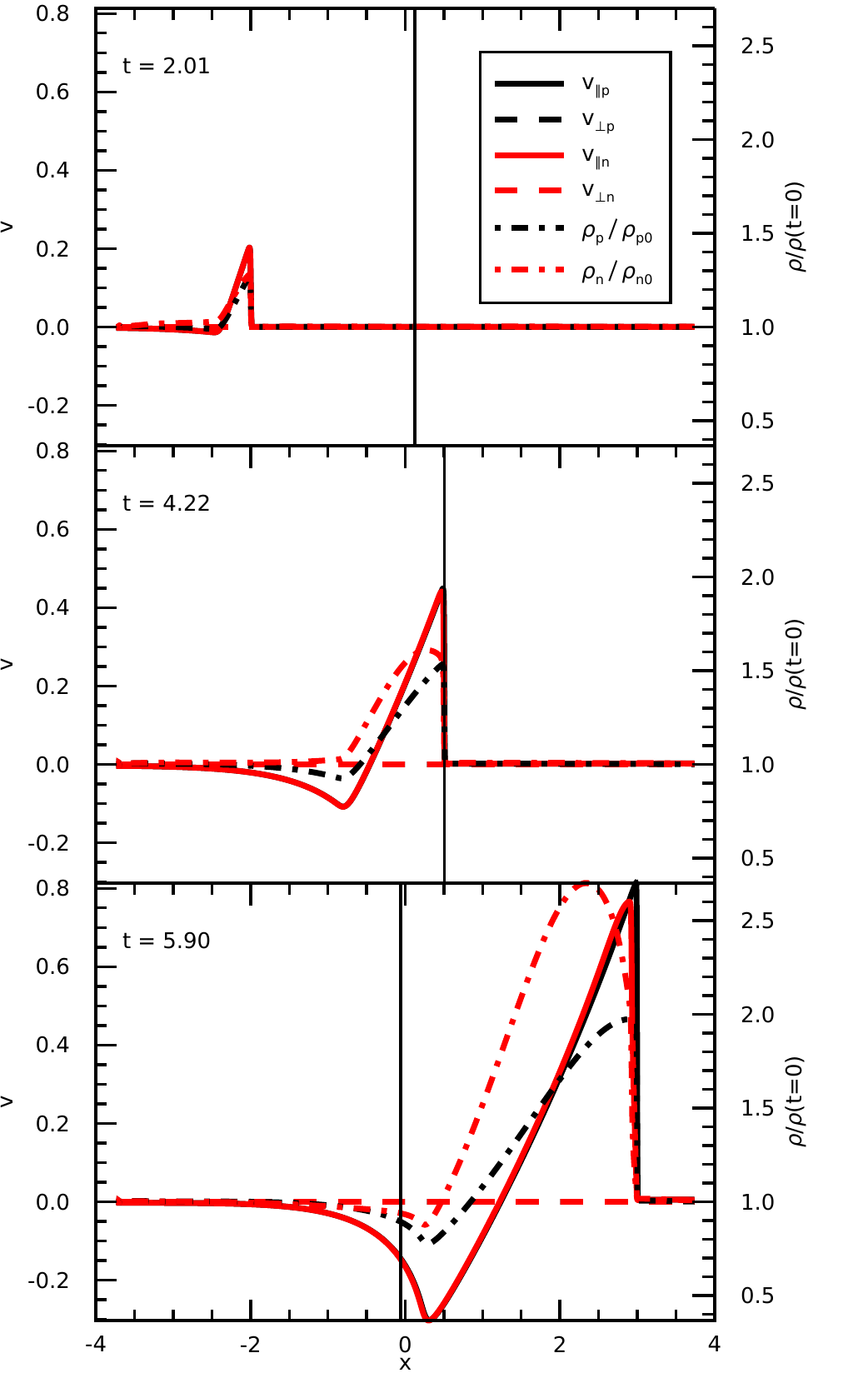}  
    \caption{MHD (left) and PIP (right) time series for a vertical field ($B_x = 1, B_y =0$), for the plasma (black) and neutral (red) species. Velocity parallel (solid) and perpendicular (dashed) is plotted and corresponds to the left axis. The density increase normalised by the initial density profile is shown as the dot-dash line and corresponds to the right axis.}
    \label{fig:straighttimeseries}
\end{figure*}

\subsection{Wave steepening} \label{sec:wavesteep}


Waves naturally steepen as they propagate through a stratified atmosphere, provided that the non-linear terms are greater than the dissipative effects. The rate at which this occurs depends on the scale height of the system. Figure \ref{fig:wavesteepen} shows a wave steepening in our atmosphere for the MHD (solid) and plasma species in the PIP (dashed) cases at different snapshots. Here the PIP case has a neutral fraction of $\xi_n = 0.9$ at the lower boundary and the same driver amplitude, hence more applied kinetic energy since there is also velocity applied to the neutral species. 

One can see from Figure \ref{fig:wavesteepen} that the amplitude of the waves is approximately equal after $0.5 t$. As time advances, the amplitude increases much more rapidly for the PIP case, with the velocity being around $20\%$ larger at $4.5 t$. The increased rate of steepening can be explained using by the pressure scale heights. In the MHD case, the pressure scale height is constant away from the boundaries (between $-3 \leq x \leq 3$). For the PIP case, the pressure scale height for a coupled wave changes as it propagates from a mostly neutral to a mostly plasma medium. 

It is also clear in Figure \ref{fig:wavesteepen} that the PIP shock propagates slower than the MHD shock. For a vertical magnetic field, the propagation speed of an isolated plasma wave is the sound speed ($c_s$) below the $\beta _v$ point, and the slow speed (which aligned with the magnetic field is the sound speed) above the $\beta _v$ point. An isolated neutral wave will propagate at the neutral sound speed ($c_{s {\rm n}}$) through all heights in the atmosphere, which is a factor of $\sqrt{2}$ smaller than the plasma sound speed, see Figure \ref{fig:wavespeeds}. Here, the coupling between the neutral and plasma species results in a propagation speed slower than for an isolated plasma (i.e., MHD), and faster than an isolated neutral species. This is discussed in more detail in later sections.

\subsection{Shock fundamentals} \label{sec:shock}
\subsubsection{Shock transitions}\label{transistions}

Shocks are classified by comparing the upstream and downstream normal flow velocity $v_{\perp}$ to the characteristic speeds \citep[e.g.,][]{Delmont2011}:
\begin{itemize}
\item (1) superfast: $V_f < \abs{v_\perp}$,
\item (2) subfast: $\hat{V}_A < \abs{v_\perp} < V_f $,
\item (3) superslow: $V_s < \abs{v_\perp} < \hat{V}_A$,
\item (4) subslow: $0< \abs{v_\perp} < V_s$,
\item ($\infty$) static: $v_\perp = 0$.
\end{itemize}
Defining the upstream condition $\text{u}$ and downstream condition $\text{d}$, several shocks of the form $\text{u} \rightarrow \text{d}$ are possible:
\begin{itemize}
\item $1 \rightarrow 2 $ fast shocks,
\item $3 \rightarrow 4$ slow shocks,
\item $1 \rightarrow 2=3$ switch-on,
\item $2=3 \rightarrow 4$ switch-off,
\item $1\rightarrow 3, 1 \rightarrow 4, 2 \rightarrow 3, 2 \rightarrow 4$ intermediate shocks.
\end{itemize}

In the neutral species, there is only one characteristic wave speed, the sound speed $c_s$, and hence only a sonic shock can exist in this fluid. A purely hydrodynamic shock in our setup will only exist in the parallel component of velocity.

In partially ionised plasma, both neutral and plasma species exist and can interact through collisional coupling. Previous studies have shown that this can have several interesting effects, including the formation of shock substructure and additional shock transitions \citep{Hillier2016,Snow2019}.

\subsubsection{Shock frame} \label{sec:shockframe}

To calculate the shock transitions accurately, the shock must be moved to the shock frame, where it is stationary in time. The shock frame here is defined as:
\begin{equation}
    x_s = (x - x_{shock})/t,
\end{equation}
where $x_{shock}$ is the location of the shock determined by the maximum absolute velocity gradient (if discontinuous), or the maximum absolute velocity (if smooth). The shock velocity is calculated using a quadratic derivative of the shock location in time.

\section{Results}
In this section we present the analysis of the compressional waves as they propagate through the stratified atmosphere with an inclined magnetic field.
The velocity from our simulations is decomposed into parallel and perpendicular components relative to the instantaneous magnetic field. 
This is approximately a decomposition into slow (parallel) and fast (perpendicular) wave modes above the $\beta _v$ point (i.e., $x>0$ in our model). Below the $\beta _v$ point ($x<0$) this approximation does not hold and only an acoustic shock exists. 

\subsection{Vertical Magnetic Field: $B_x=1.0, B_y=0.0$} \label{sec:straight}

\begin{figure*} 
    \centering
    \includegraphics[width=0.99\textwidth,clip=true,trim=0.91cm 10.85cm 0.8cm 11.1cm]{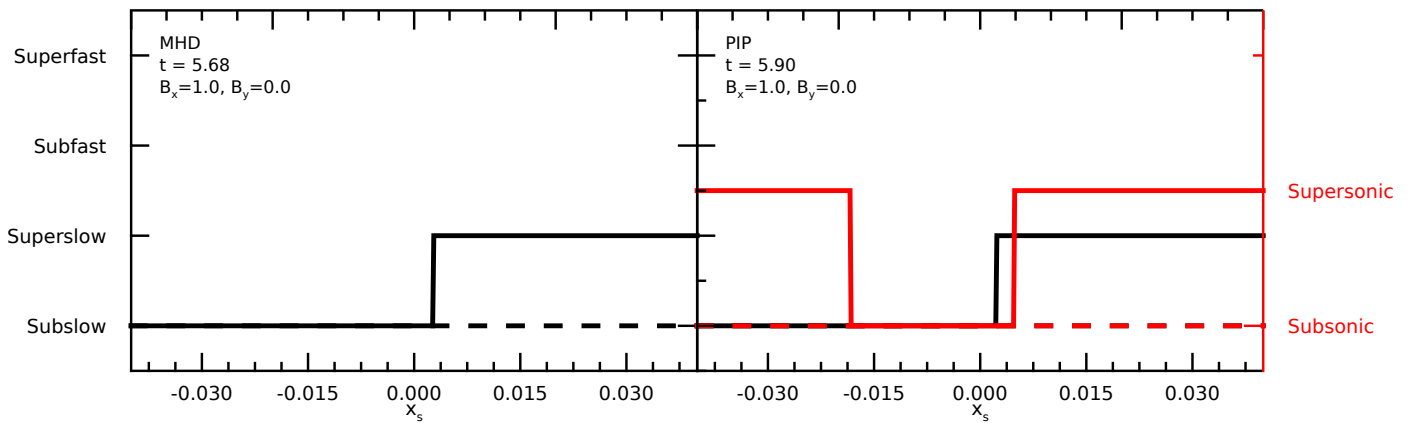}
\caption{MHD shock transitions for the vertical magnetic field ($B_x=1,B_y=0$) at $t = 5.68$ (left). Horizontal axis has been transferred to the shock frame $x_s$. PIP shock transitions for the vertical magnetic field ($B-x=1,B_y=0$) for the parallel shock transitions for the plasma (black) and neutral (red) species at $t=5.90$ (right).}
    \label{fig:MHSstraight}
\end{figure*}

\begin{figure} 
    \centering
    \includegraphics[width=0.49\textwidth,clip=true,trim=1.9cm 8cm 2cm 9cm]{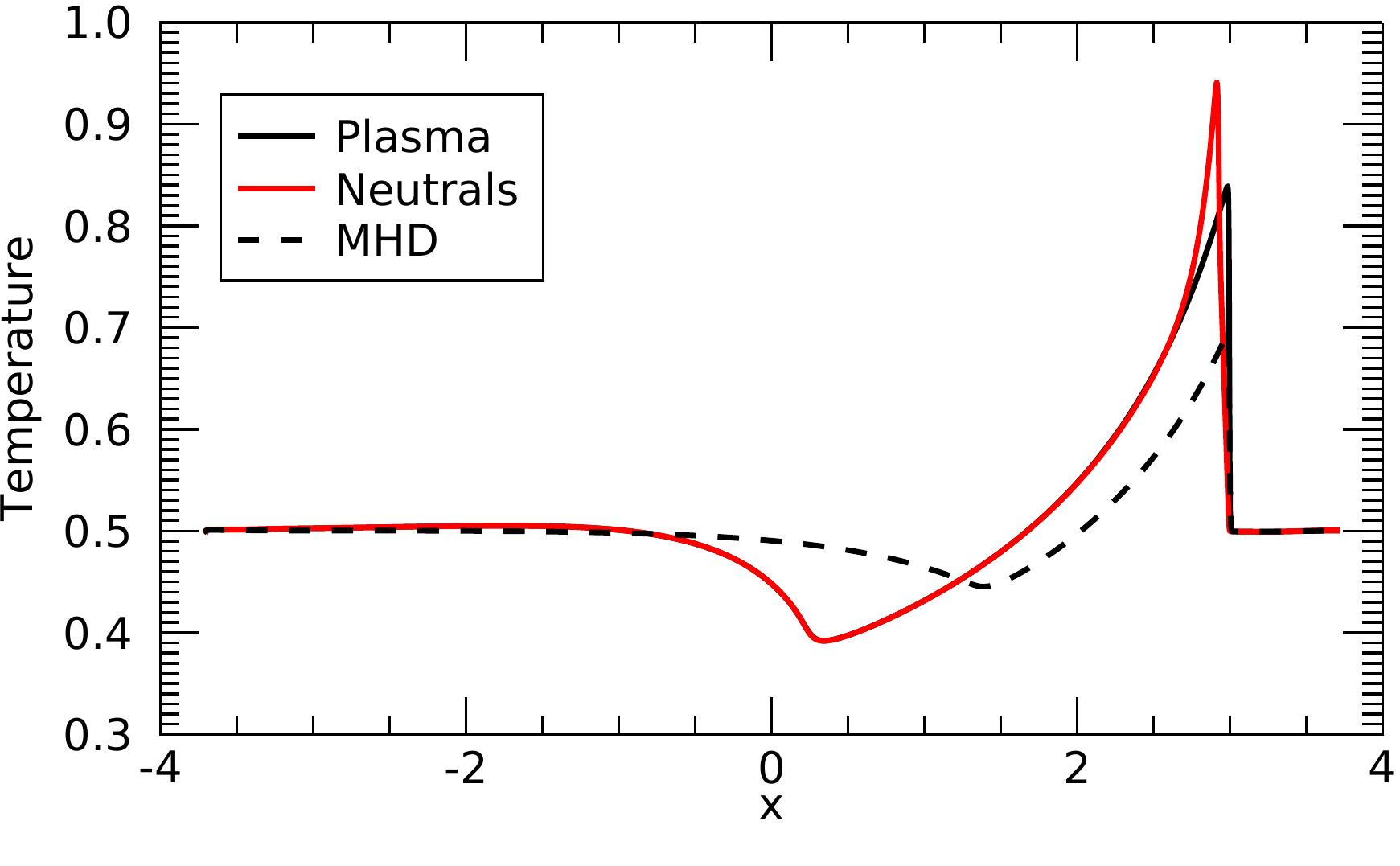} \\
    \includegraphics[width=0.49\textwidth,clip=true,trim=0.8cm 8cm 2.2cm 8.5cm]{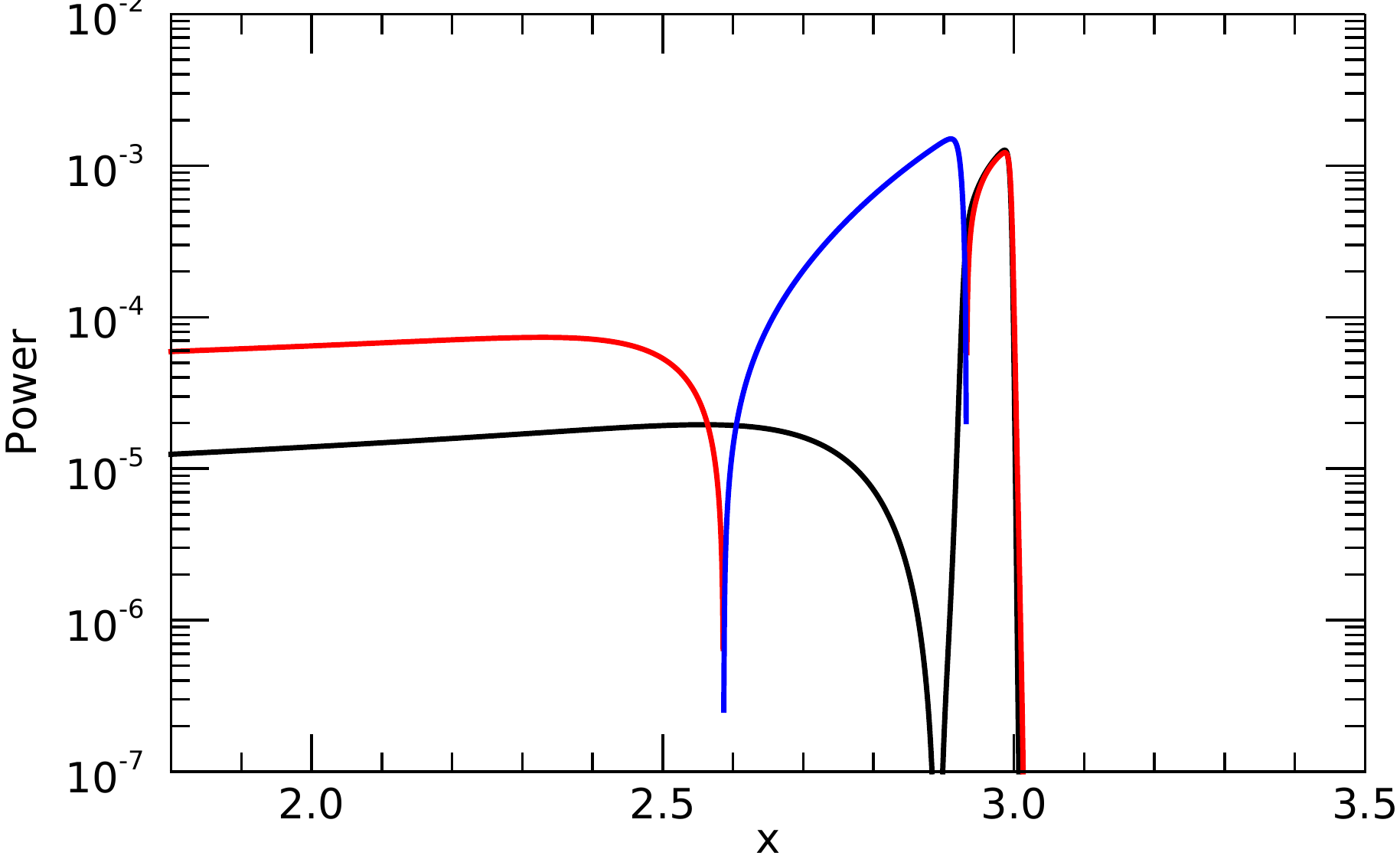}
    \caption{(Top) temperature in the PIP simulation for the plasma (black) and neutral (red) species. The black dashed line shows the MHD temperature at the same time. (Lower) Frictional heating (black) and thermal damping (red and blue). Red line indicates ions losing heat to neutrals. Vice versa for the blue line.}
    \label{fig:PIPstraightheating}
\end{figure}

\subsubsection{MHD}
For a vertical magnetic field ($B_x=1, B_y=0$) the wave steepens into a shock below the $\beta _v$ point. A vertical magnetic field acts as a reference case since the propagation is restricted to be along the magnetic field direction and $v_{\perp}=0$. The wave steepens with height, forming a shock below the $\beta _v$ point, see Figure \ref{fig:straighttimeseries}. The shock front is characterised by the jump in the parallel velocity and density.

Figure \ref{fig:MHSstraight} shows the shock transitions corresponding to the last frame of the time series (Figure \ref{fig:straighttimeseries}), where the horizontal axis has been shifted into the shock frame (as described in Section \ref{sec:shock}). The shock transitions show that this is a slow-mode shock (i.e., a transition from above to below the slow speed). This is as expected for this atmosphere, as above the $\beta_v$ point only the slow mode transition parallel to the field is permitted.

\subsubsection{PIP}

\begin{figure}
    \centering
    \includegraphics[width=0.5\textwidth,clip=true,trim=1.7cm 8.5cm 2cm 9cm]{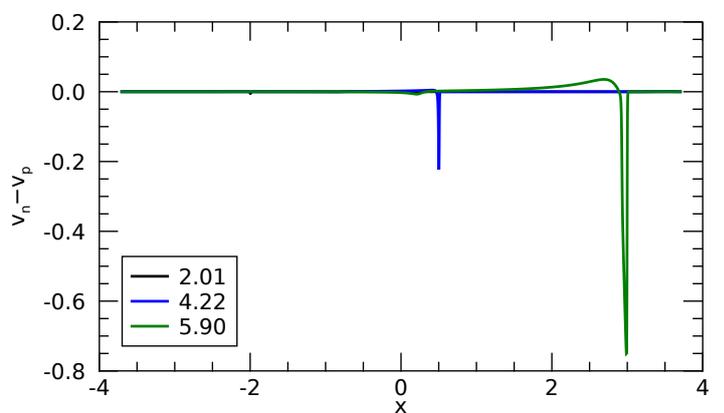}
    \caption{Drift velocities at different times for the vertical magnetic field ($B_x = 1.0, B_y=0.0$) at different times, corresponding to the snapshots in Figure \ref{fig:straighttimeseries}.}
    \label{fig:driftvelsstraight}
\end{figure}

For the PIP case, the overall shock behaviour is similar to the MHD case however there are interesting features localised around the shock front, Figure \ref{fig:straighttimeseries}. The wave steepens into a shock below the $\beta _v$ line showing sharp discontinuities in both the plasma (black) and neutral (red) species. The perpendicular component of both species is zero, as expected. A sharp rise in density is present at the shock front in both species. The neutral density increases by a larger relative amount than the plasma, however the neutral density is very small here (see Figure \ref{fig:densities}). 

There is a slight lag between the maximum velocities in the two species, with the neutral fluid slightly behind the plasma. Physically, the plasma sound speed $c_{s}$ is larger than the neutral sound speed $c_{sn}$, see Figure \ref{fig:wavespeeds}. For isolated species, a plasma shock would be expected to propagate faster in this atmosphere (see Figure \ref{fig:wavesteepen}). High in the atmosphere, the neutral fraction is very small and hence the collisional coupling is weaker.  As such, the neutral species is dragged by the plasma via collisional coupling. 

The shock transitions (Figure \ref{fig:MHSstraight}) for the PIP case show a show mode shock in the plasma and a sonic shock in the neutral species. Note that the shock transitions are only applicable around $x_s =0$ and the transition at $x_s\approx -0.017$ is an artefact from the limitations of the shock fit and not a real shock transition.

Figure \ref{fig:PIPstraightheating} shows the temperature across the domain (a) and the heating terms (b), where the frictional heating is defined as $\alpha _c (T_{\text{n}},T_{\text{p}}) \rho_{\text{n}} \rho_{\text{p}} (v_{\text{n}}-v_{\text{p}})^2/2$ and thermal damping is $3 \alpha _c (T_{\text{n}},T_{\text{p}}) \rho_{\text{n}} \rho_{\text{p}} (P_{\text{n}}/\rho_{\text{n}}-P_{\text{p}}/(2\rho_{\text{p}}))/2$. 
The shock causes a temperature increase in both plasma and neutral species, however the temperature in the neutrals increase comparatively more. 
Two reasons exist for the higher neutral temperature: the neutrals undergo greater compression because the wave is dragged faster than it would travel in a single fluid simulation, and the lower neutral density at this point which allows for a greater temperature increase from the frictional heating.

The MHD temperature at the same time is also shown in Figure \ref{fig:PIPstraightheating} (black dashed line). 
The PIP simulation has a higher temperature than the MHD case. This is likely to be as a result of the increased nonlinearity of the wave created by the smaller pressure scale height of the neutrals.

The drift velocities ($v_{n}-v_{p}$) are shown for a few different times in Figure \ref{fig:driftvelsstraight}. 
As the shock propagates, the drift velocities increase. At time $\tau =5.50$ the drift velocity is fairly close to the equilibrium neutral sound speed ($1/\sqrt{2}$). 
It should be noted that such large drift velocities are approaching the limits of applicability of our drift velocity model (Equation \ref{drift}), because once the drift velocity becomes much greater than the sound speed it is the drift and not thermal motions that drive the collisions between the fluids \citep{DRAINE1986}

\subsection{Inclined Field: $B_x=0.8,B_y=0.6$} \label{sec:a1}

\begin{figure*} 
    \centering
    \includegraphics[width=0.49\textwidth,clip=true,trim=5.85cm 8cm 5.15cm 3.2cm]{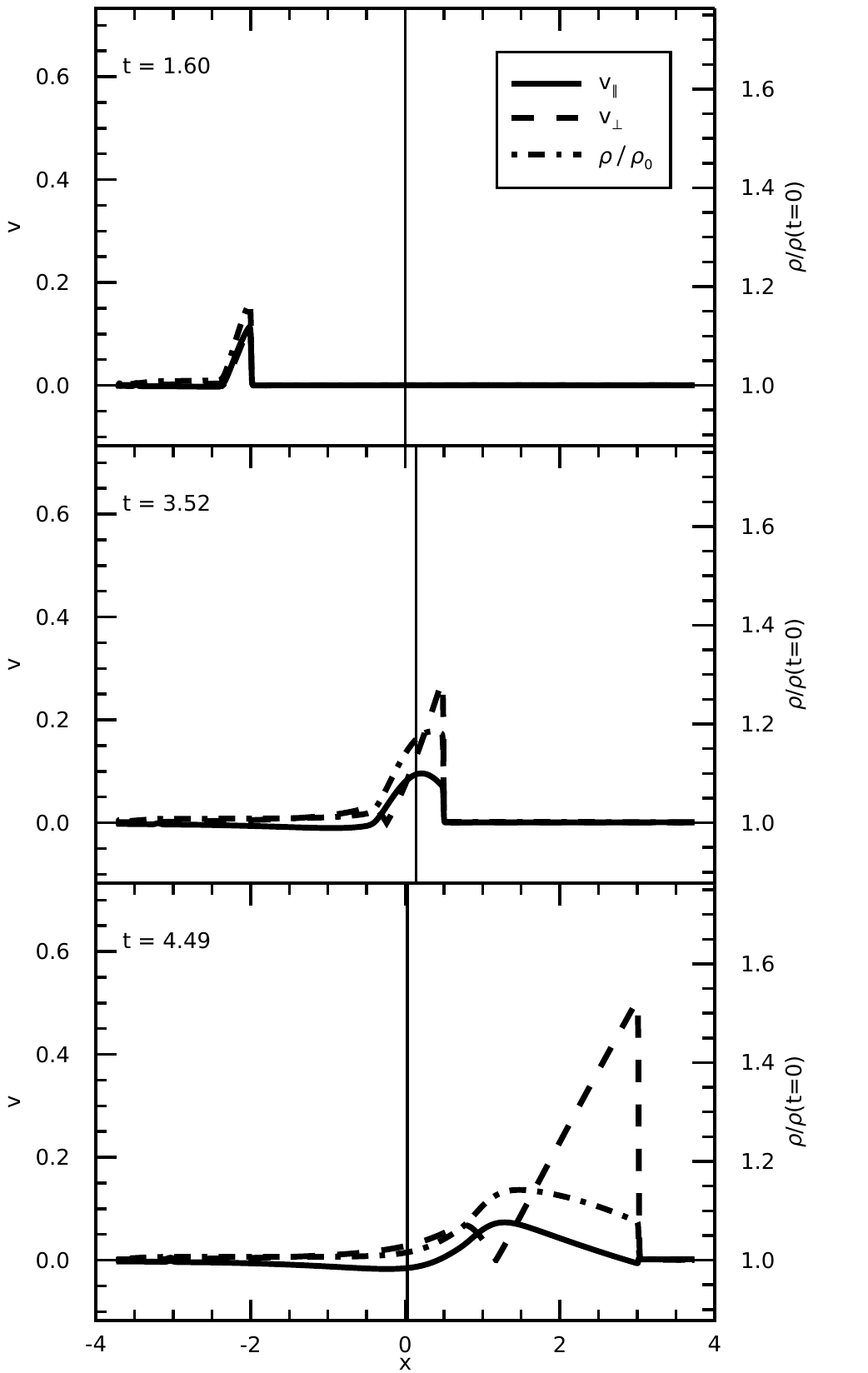}
    \includegraphics[width=0.49\textwidth,clip=true,trim=5.85cm 8cm 5.15cm 3.2cm]{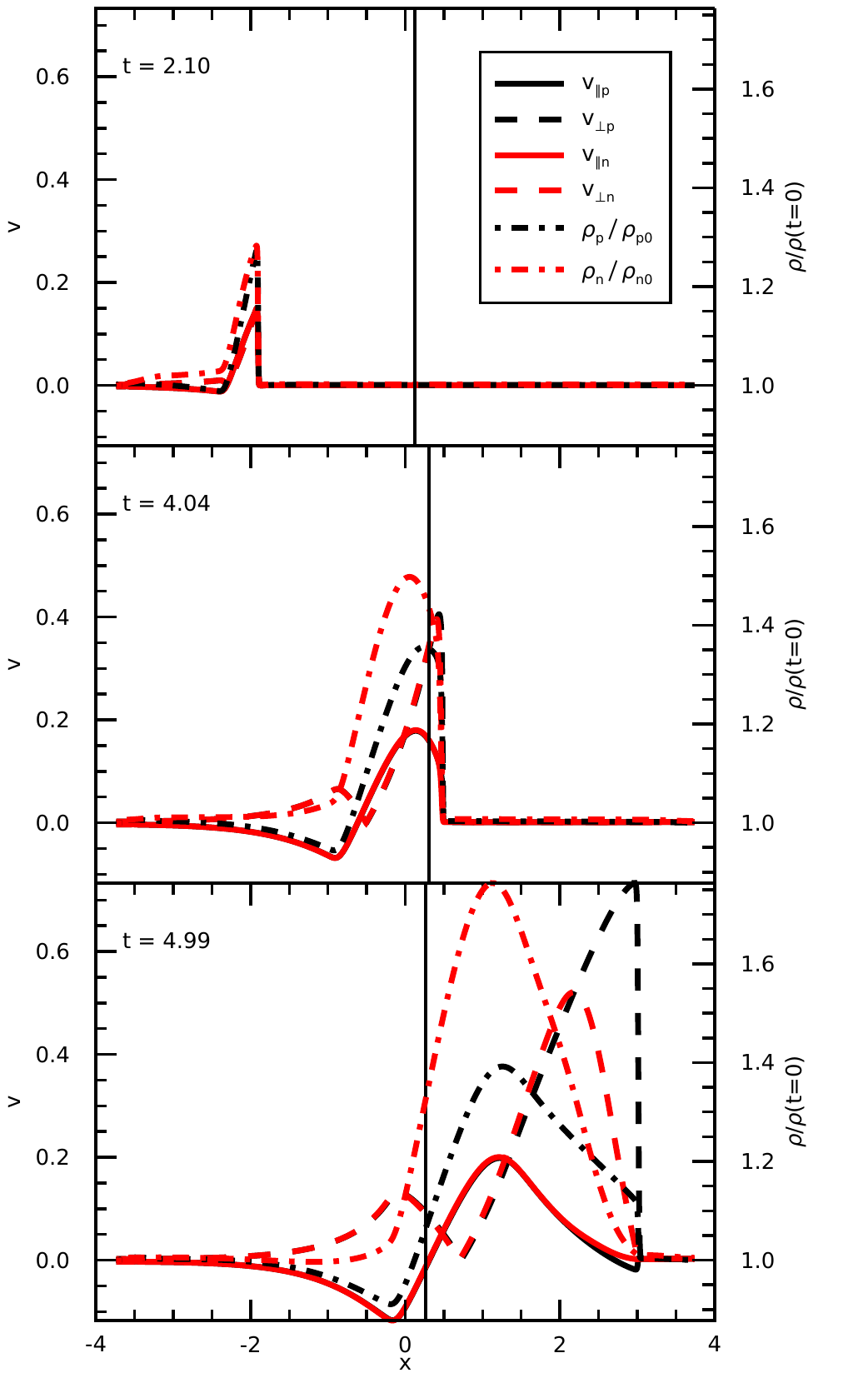}
    \caption{MHD (left) and PIP(right) time series for an inclined field with $B_x=0.8,B_y=0.6$ for the plasma (black) and neutral (red) species. Solid (dashed) line shows the parallel (perpendicular) component of velocity relative to the instantaneous magnetic field. Density normalised by the density at $t=0$ is plotted as the dot-dashed line and corresponds to the right axis.}
    \label{fig:bothtimeseriesa1}
\end{figure*}

\begin{figure*}
    \centering
    \includegraphics[width=0.99\textwidth,clip=true,trim=0.91cm 10.85cm 0.8cm 11.1cm]{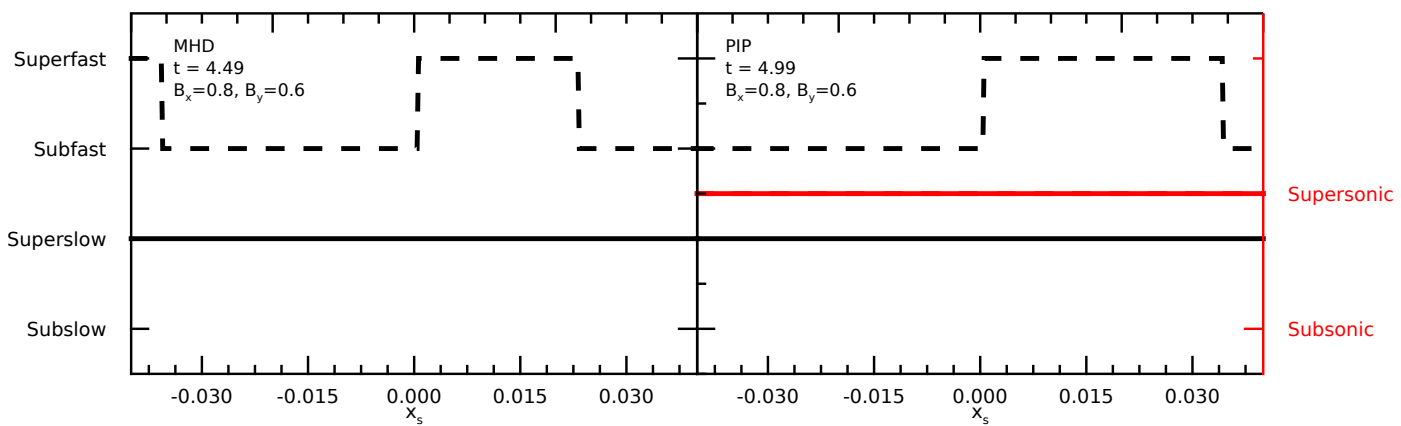}
    \caption{Shock transitions for the MHD (left panel) and PIP (right panel) shocks with an inclined field ($B_x=0.8,B_y=0.6$). Parallel (solid) and perpendicular (dashed) components are plotted in their relative shock frames. The neutral transitions (red) correspond to the right axis.}
    \label{fig:bothangletrans}
\end{figure*}

\begin{figure}
    \centering
    \includegraphics[width=0.49\textwidth,clip=true,trim=1.6cm 8cm 2cm 9cm]{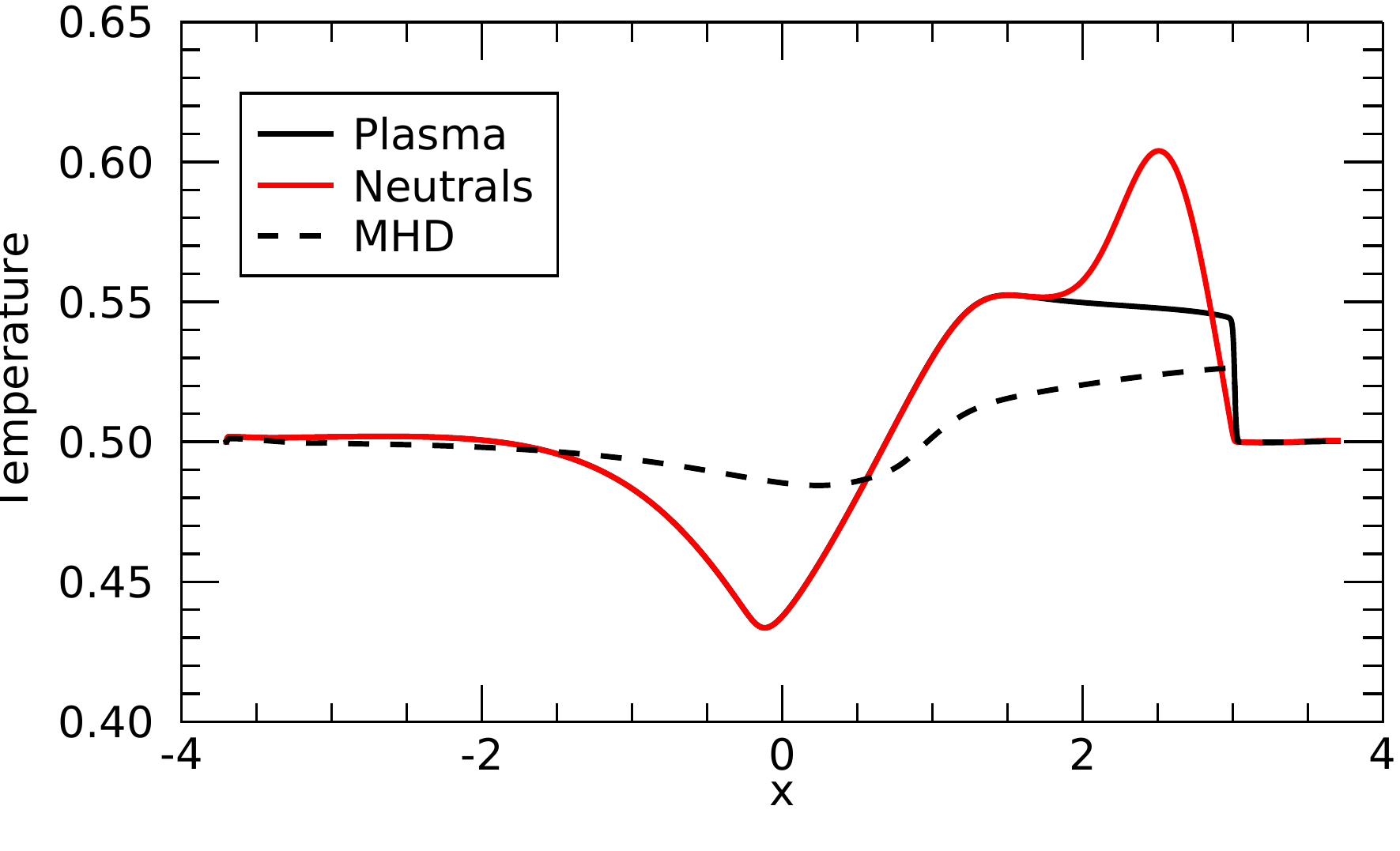} \\
    \includegraphics[width=0.49\textwidth,clip=true,trim=0.8cm 8cm 2.2cm 8.4cm]{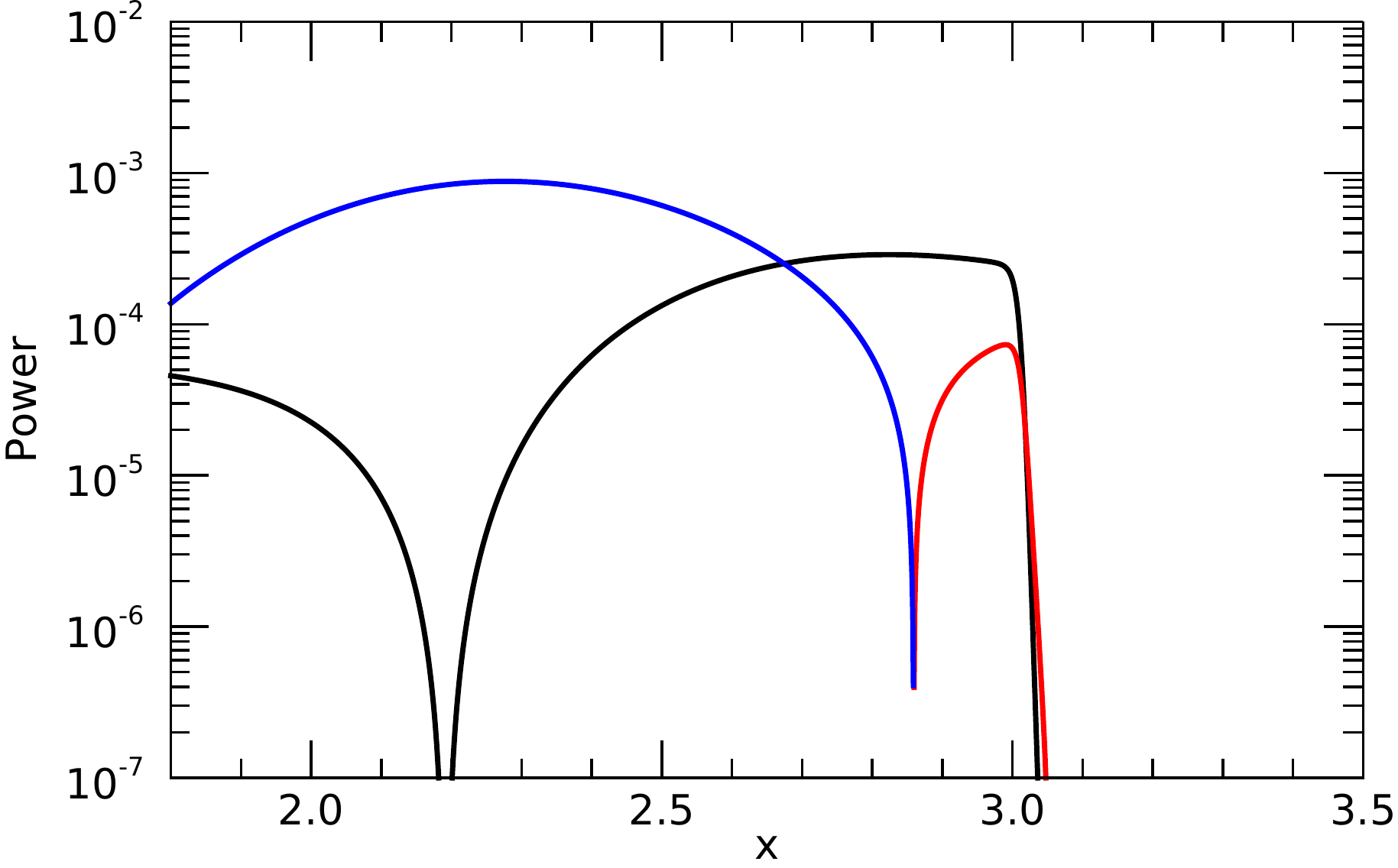}
    \caption{Inclined field: (left) temperature in the PIP simulation for the plasma (black) and neutral (red) species, and the MHD case (black dashed) Note that the MHD is at $\tau = 4.31$ and the PIP simulation is at time $\tau = 4.80$. (right) Frictional heating (black) and thermal damping (red and blue). Red line indicates ions losing heat to neutrals. Vice versa for the blue line.}
    \label{fig:PIPheatinga1}
\end{figure}

When the initial conditions feature an angle in the magnetic field, energy can be exchanged between the different wave modes at the $\beta _v$ point and the shock fronts can separate into their fast (perpendicular) and slow (parallel) components \citep[studied in MHD by][]{Pennicott2019}. Here we present results from a inclined magnetic field with $B_x =0.8, B_y=0.6$, preserving the total magnetic field strength as unity. The initial pressure and density profiles are identical to the vertical field (see Figure \ref{fig:densities}), as is the driver. 


\subsubsection{MHD case ($B_x=0.8,B_y=0.6$)} \label{sec:a1MHD}
In the MHD case (see left column of Figure \ref{fig:bothtimeseriesa1}), a notable difference (compared to the vertical field) is the non-zero component of velocity perpendicular to the magnetic field (i.e., the fast-mode component). 
Below the $\beta_v$ point, there is a single shock front that consists of both perpendicular and parallel components of velocity, see Figure \ref{fig:bothtimeseriesa1}. As the shock propagates through the $\beta _v$ point, the energy is exchanged between modes and the fast- and slow-mode separate. 
Notably, the parallel (slow) component of the velocity becomes smoothed: below $x=0$ there is a steep discontinuity, above $x=0$ the wavefront becomes smoothed and no longer satisfies the conditions to be called a shock. 
The perpendicular (fast) component retains its steep shock structure and as a consequence of being a shock there is a steep change in the density relative to the initial density (Figure \ref{fig:bothtimeseriesa1}). 
A small non-zero parallel velocity component exists at the location of the fast-mode shock which is an artefact from the assumption of decomposing into parallel and perpendicular velocities, \citep[see, for example,][]{Pennicott2019}.

The shock transitions have been calculated for the wide inclined field, $B_x=0.8,B_y=0.6$, and are shown in Figure \ref{fig:bothangletrans}. In this case, there is no shock transition associated with the parallel component of velocity. In the perpendicular component of velocity, there is a clear transition from superfast to subfast, i.e., a fast-mode shock. Note that shock frame here was calculated using the maximum absolute gradient for the perpendicular velocity, and the maximum absolute velocity for the parallel component, see Section \ref{sec:shockframe}. 

\subsubsection{PIP case ($B_x=0.8,B_y=0.6$)} \label{sec:a1PIP}

Below the $\beta_v$ point, the plasma and neutral species are well coupled, as are the parallel and perpendicular components of each species, see Figure \ref{fig:bothtimeseriesa1}. Changes arise above the $\beta _v$ point. The plasma components separates into its parallel (slow) and perpendicular (fast) components, with the parallel components becoming smoothed and the perpendicular component remaining sharp (as seen in the MHD case). The shock transitions (Figure \ref{fig:bothangletrans}) reveal a fast-mode shock in the perpendicular component of the plasma species, with no shock transitions present in the parallel plasma velocity or the neutral species. 

The neutral species exhibits a non-zero perpendicular velocity that propagates with (but behind) the plasma perpendicular velocity. As the neutral species does not have an associated fast speed, the perpendicular component of the neutral species arises from collisional coupling with the plasma species. 
This results in a reasonably large increase in the neutral temperature (compared to the plasma temperature) around the fast shock, Figure \ref{fig:PIPheatinga1}. 
Here, the increase in neutral temperature is at the location of the fast-mode shock only. The slow component has been smoothed and the neutral and plasma species are reasonably well coupled, unlike the vertical field where the species decouple at the shock front and there is an associated increase in frictional heating of the neutrals (see Figures \ref{fig:straighttimeseries} and \ref{fig:PIPstraightheating}). 

\begin{figure}
    \centering
    \includegraphics[width=0.49\textwidth,clip=true,trim=1.7cm 8.5cm 2cm 9cm]{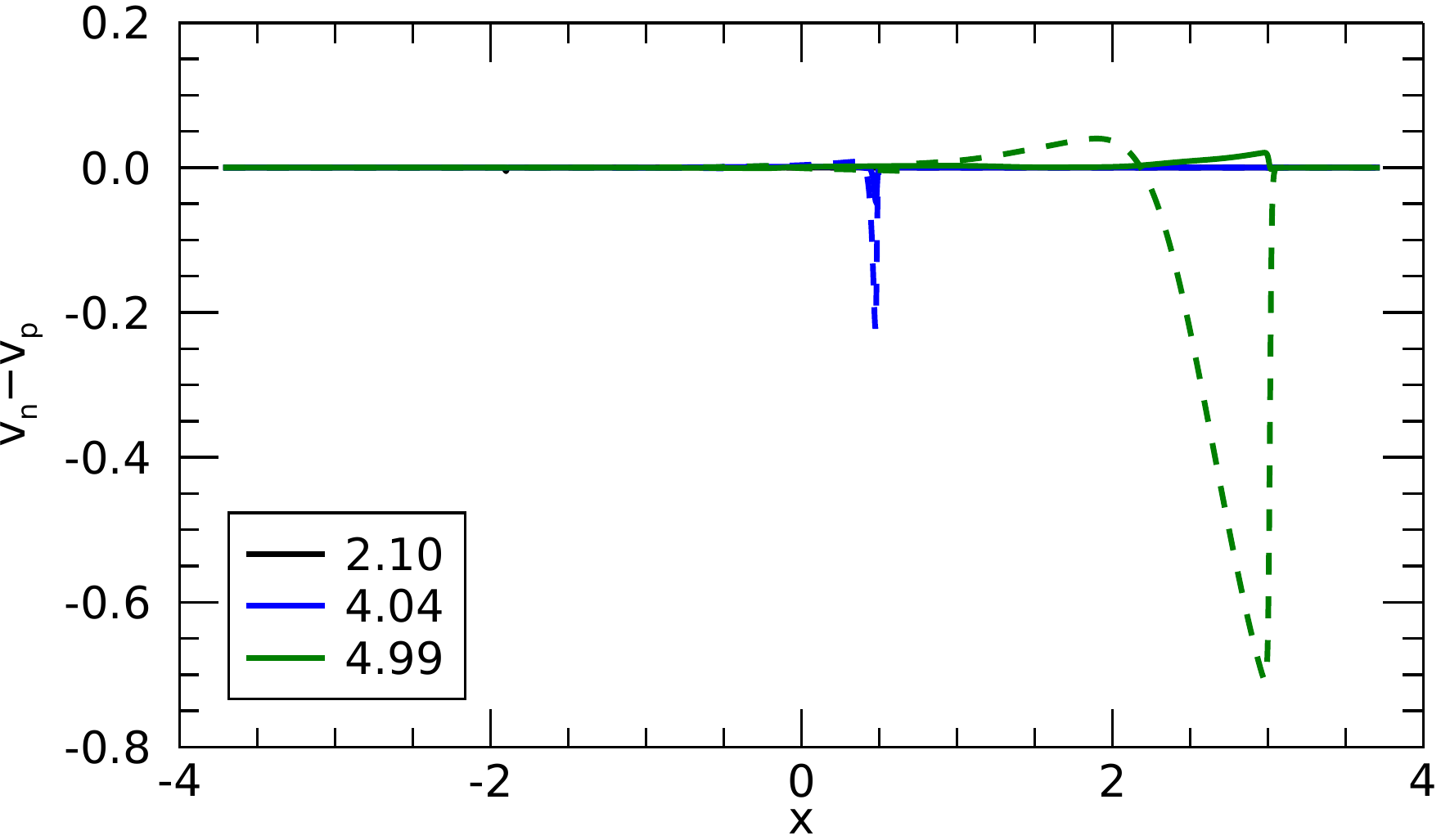}
    \caption{Parallel (solid) and perpendicular (dashed) drift velocities for the wide inclined field ($B_x=0.8, B_y=0.6$) at different times, corresponding to the snapshots in Figure \ref{fig:bothtimeseriesa1}.}
    \label{fig:driftvelsa1}
\end{figure}

The drift velocities ($v_n - v_p$) are shown in Figure \ref{fig:driftvelsa1} for a few different times. The drift velocity is fairly large at the location of the fast-mode shock. The drift velocity associated with the parallel component of velocity is relatively small. For a linear wave, coupling is much more effective than a non-linear wave. The parallel component is smoothed and no longer satisfies the conditions for a shock above the $\beta _v$ point, and the plasma and neutral species are reasonably well coupled. 

\section{Effect of different neutral fractions and collisional coupling} \label{sec:colleffect}

The cases considered thus far are using a neutral fraction of $\xi _n =0.9$ at the lower boundary with the collisional coefficient taken to be $\alpha_0 = 100$. This profile becomes plasma-dominated at $x \approx -1$, which is below the $\beta _v$ point. As such, the neutrals effect on the plasma is relatively small near the top of this domain. By increasing the initial neutral fraction, the neutral species has a larger impact on the plasma through increased collisions. A large neutral fraction, along with larger values of $\alpha_0$, is a regime that is closer to that of the solar chromosphere making it an important regime to study. In this section, larger neutral fractions and a range of $\alpha_0$ values are investigated, but first we discuss further peculiarities of this set-up.

\subsection{Different $\beta_v$ locations}

\begin{figure}
    \centering
    \includegraphics[width=0.49\textwidth,clip=true,trim=1.6cm 8.4cm 1.6cm 8.4cm]{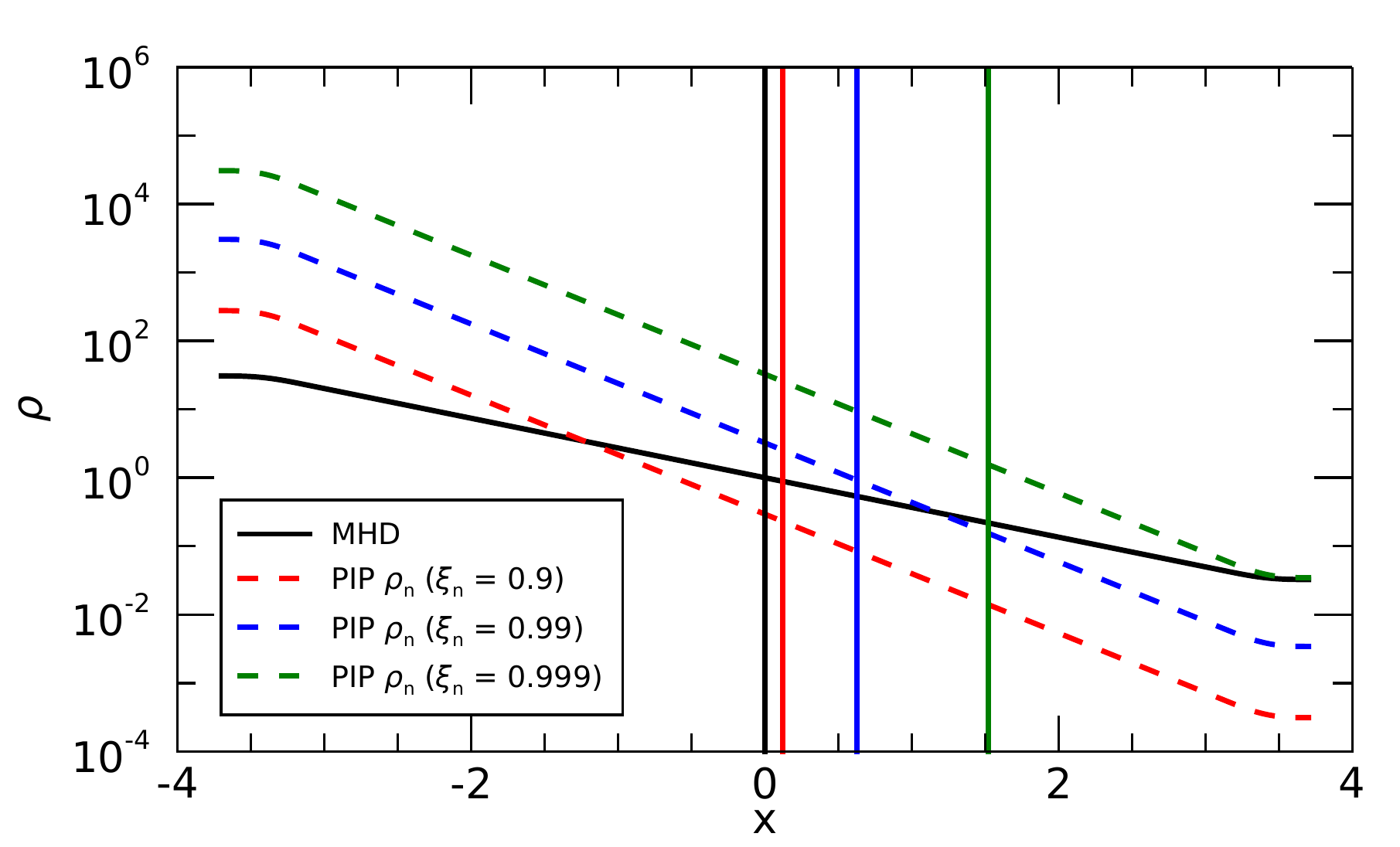}
    \caption{Plasma (solid black) and neutral (dashed) densities with height for $\xi_{n0}=0.9$ (red), $0.99$ (blue), and $0.999$ (green). The MHD density is given by the black line and the plasma component of the two-fluid simulation is identical to the MHD density profile. The vertical line indicate the $\beta _v$ point (where $c_s=v_A$) for the plasma species (and MHD) only. The coloured vertical lines show the $\beta_{vt}$ points using the total densities ($\rho_n+\rho_p$), i.e., $c_{st}=v_{At}$.}
    \label{fig:pipatmosxin}
\end{figure}

In single fluid MHD, the $\beta _v$ point is singularly defined since there is only one sound speed, and one Alfv\'en speed. In the two-fluid case, one can consider the $\beta _v$ point using isolated plasma speeds, or the coupled speeds (see Section \ref{sec:wavespeeds}) here referred to as $\beta _{vt}$ (where $c_{st}=v_{At}$). As the neutral fraction is increased, the relative locations of the $\beta_v$ and $\beta _{vt}$ points change.

Figure \ref{fig:pipatmosxin} shows the initial atmospheres for different neutral fractions specified at the base of the domain, specifically $\xi _n =0.9,0.99,0.999$. The normalisation means that the plasma density is equal to the MHD density in all atmospheres and a higher neutral fraction adds more mass to the system. The solid vertical lines show the $\beta_v$ and $\beta _{vt}$ points for all atmospheres. The black line uses the plasma speeds only ($c_s=v_A$) and hence is the same for all atmospheres here, whereas the coloured lines use the total speeds ($c_{st}=v_{At}$) and varies with the neutral fraction. For $\xi _n =0.9$, the two lines are fairly close together (red for coupled, black for MHD). For larger $\xi_n$ values, the coupled $\beta _v$ occurs much higher in the atmosphere than the plasma $\beta _v$. 

The previous sections of this paper use a neutral fraction of $\xi _n =0.9$ and there is very little difference between the $\beta_v$ and $\beta_{vt}$ points (Figure \ref{fig:pipatmosxin}). The obvious question here is: does the mode conversion occur at the $\beta _v$ ($c_s=v_A$) or the $\beta _{vt}$ ($c_{st}=v_{At}$) point? This is investigated by varying the collisional coefficient and the neutral fraction.

Similarly, care must be taken when defining the wave speeds; for a weakly coupled case, the plasma wave speeds ($c_s, v_A, V_s, V_f$) are likely most appropriate, whereas for a strongly coupled case the combined wave speeds ($c_{st}, v_{At}, V_{st}, V_{ft}$) should be more suitable, see Section \ref{sec:wavespeeds}. For finite coupling, it is not immediately obvious which wave speeds are most appropriate. For the case where $\xi_n=0.9$ at the base of the domain, the neutral density and pressure are relatively small compared to the plasma values above the $\beta _v$ point, hence there is not too much of a difference. Choosing the appropriate wave speeds becomes more important when the neutral fraction is higher, see Figure \ref{fig:pipatmosxin} where the neutral and plasma densities are approximately equal near the top of the domain.


\subsection{Effect of different collisional coefficients} \label{sec:increasedcoll}

\begin{figure}
    \centering
    \includegraphics[width=0.49\textwidth,clip=true,trim=1.8cm 8.4cm 1.6cm 8.4cm]{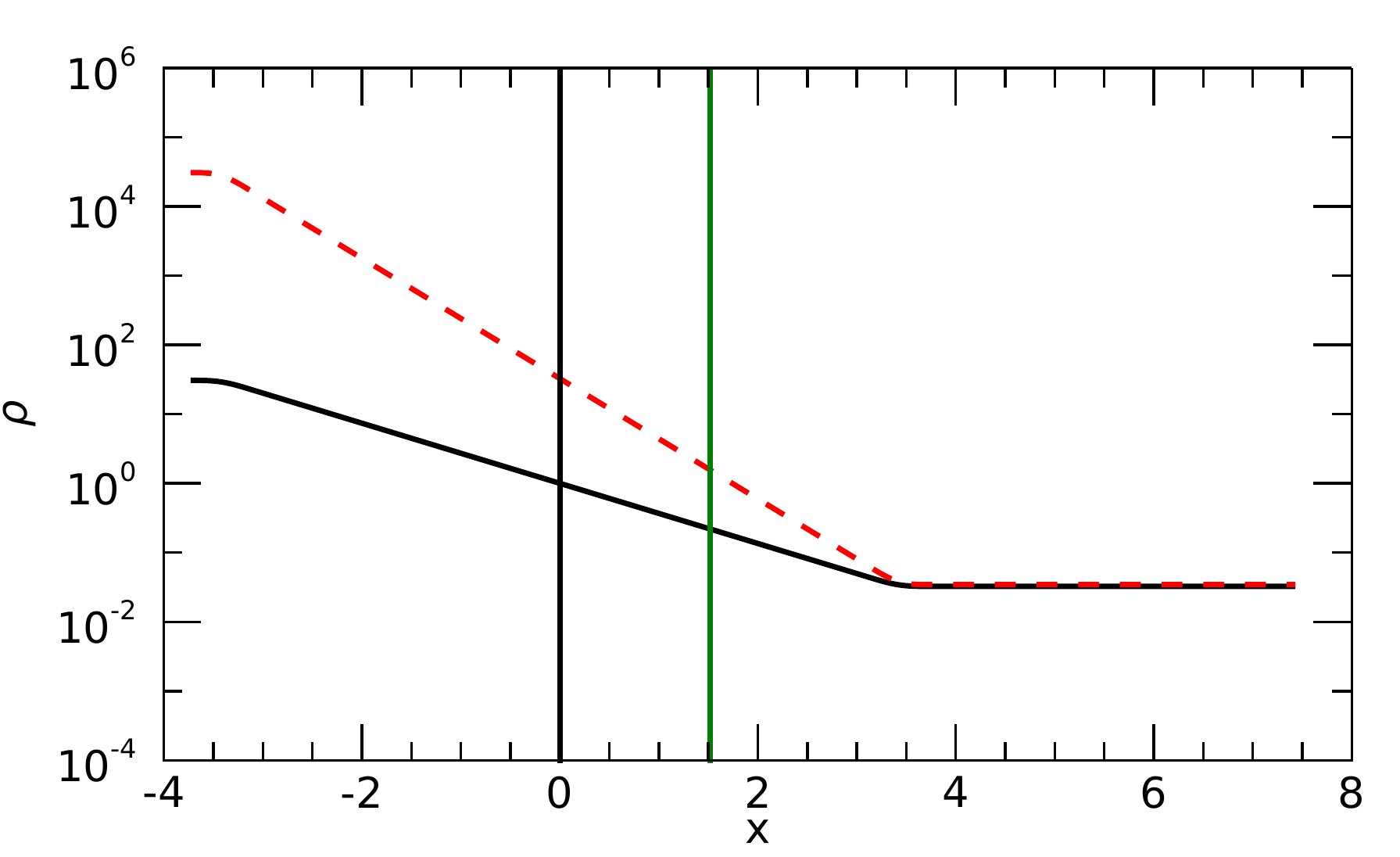}
    \caption{Long atmosphere used for the collisional coefficient simulations. Black line shows the plasma (and MHD) density. Red dashed line shows the neutral density for an initial neutral fraction of $\xi _n =0.999$. The $\beta _v$ and $\beta _{vt}$ points are shown by the black and green vertical lines.}
    \label{fig:longatmos}
\end{figure}

\begin{figure}
    \centering
    \includegraphics[width=0.49\textwidth,clip=true,trim=5.8cm 2.71cm 6.4cm 1.4cm]{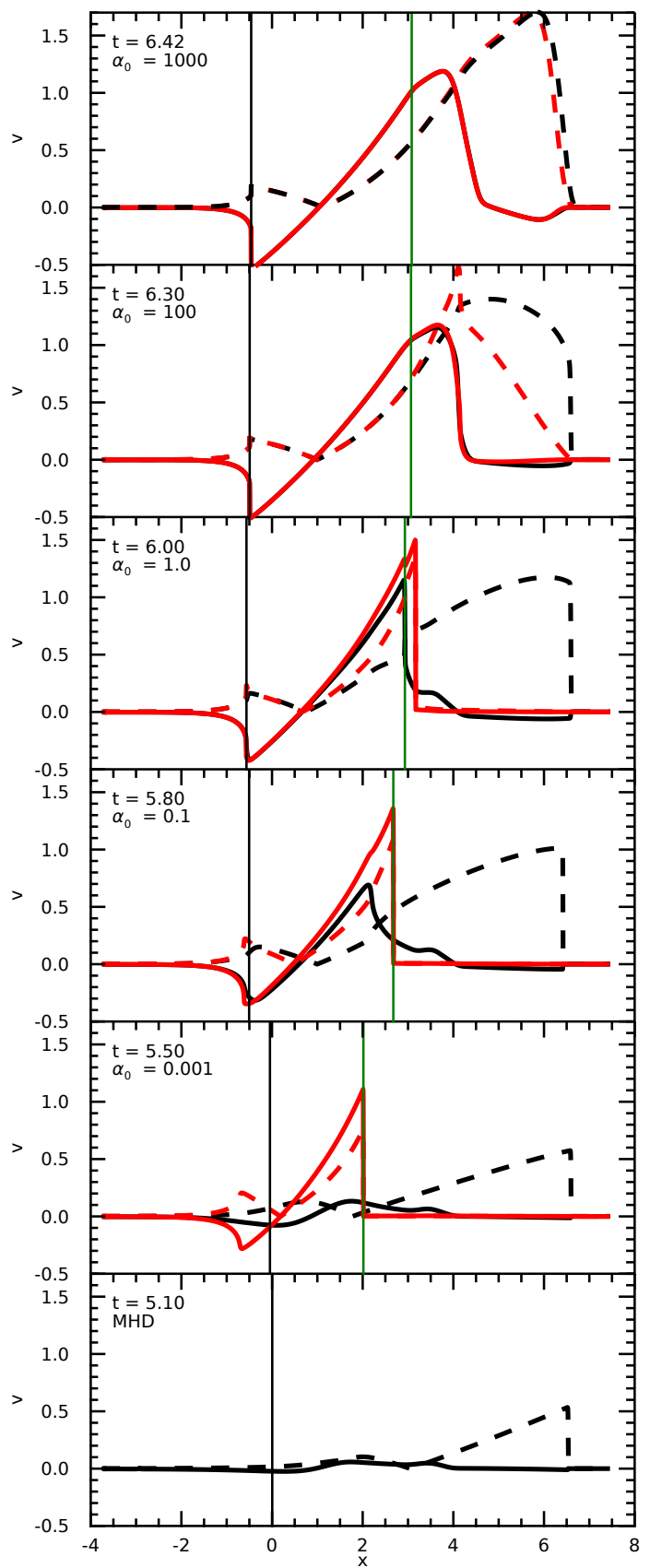}
    \caption{Snapshots of the parallel (solid) and perpendicular (dashed) velocities for the plasma (black) and the neutral (red) species for different collisional coefficients.}
    \label{fig:alpsnap}
\end{figure}

\begin{figure}
    \centering
    \includegraphics[width=0.49\textwidth,clip=true,trim=1.7cm 8.2cm 1.85cm 8.4cm]{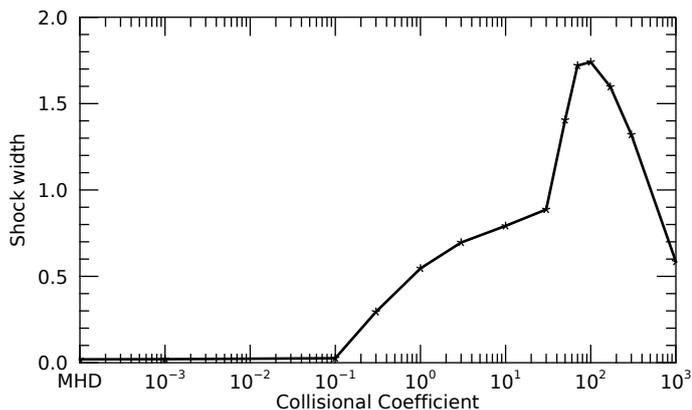}
    \caption{Perpendicular shock widths for different collisional coefficients. Corresponds to the snapshots in Figure \protect\ref{fig:alpsnap}. Shock width is calculated as the distance between the maximum velocity, and the maximum absolute gradient of velocity.}
    \label{fig:alpsnapwidth}
\end{figure}

\begin{figure}
    \centering
    \includegraphics[width=0.495\textwidth,clip=true,trim=5.2cm 2.5cm 5.4cm 1.4cm]{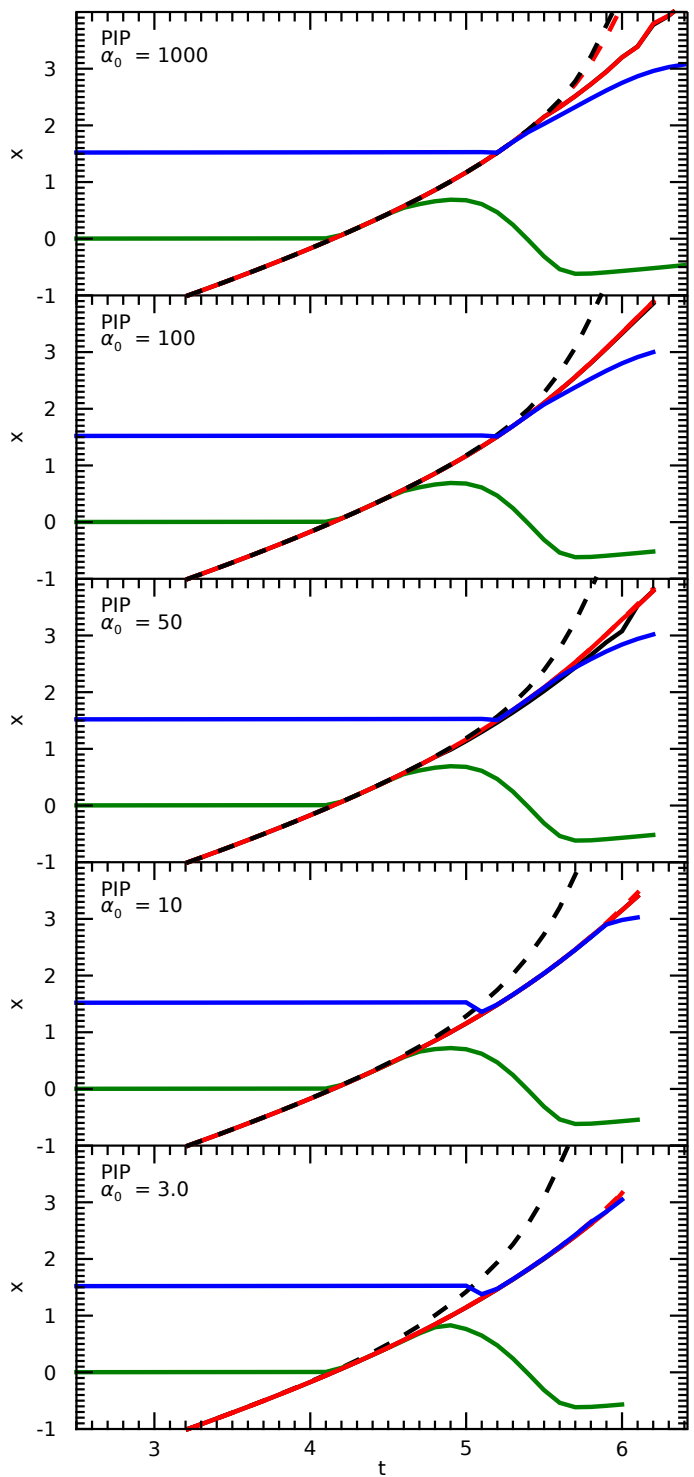}
    \caption{Shock locations through time for different collisional coupling coefficients. Parallel (solid) and perpendicular (dashed) velocities and shown for the plasma (black) and neutral (red) species. The $\beta _v$ and $\beta _{vt}$ locations are shown as the green and blue line respectively. The graph is plotted until the perpendicular shock approaches the top of the domain.}
    \label{fig:shocklocall}
\end{figure}

\begin{figure*}
    \centering
    \includegraphics[width=0.99\textwidth,clip=true,trim=1.6cm 10cm 1.4cm 10cm]{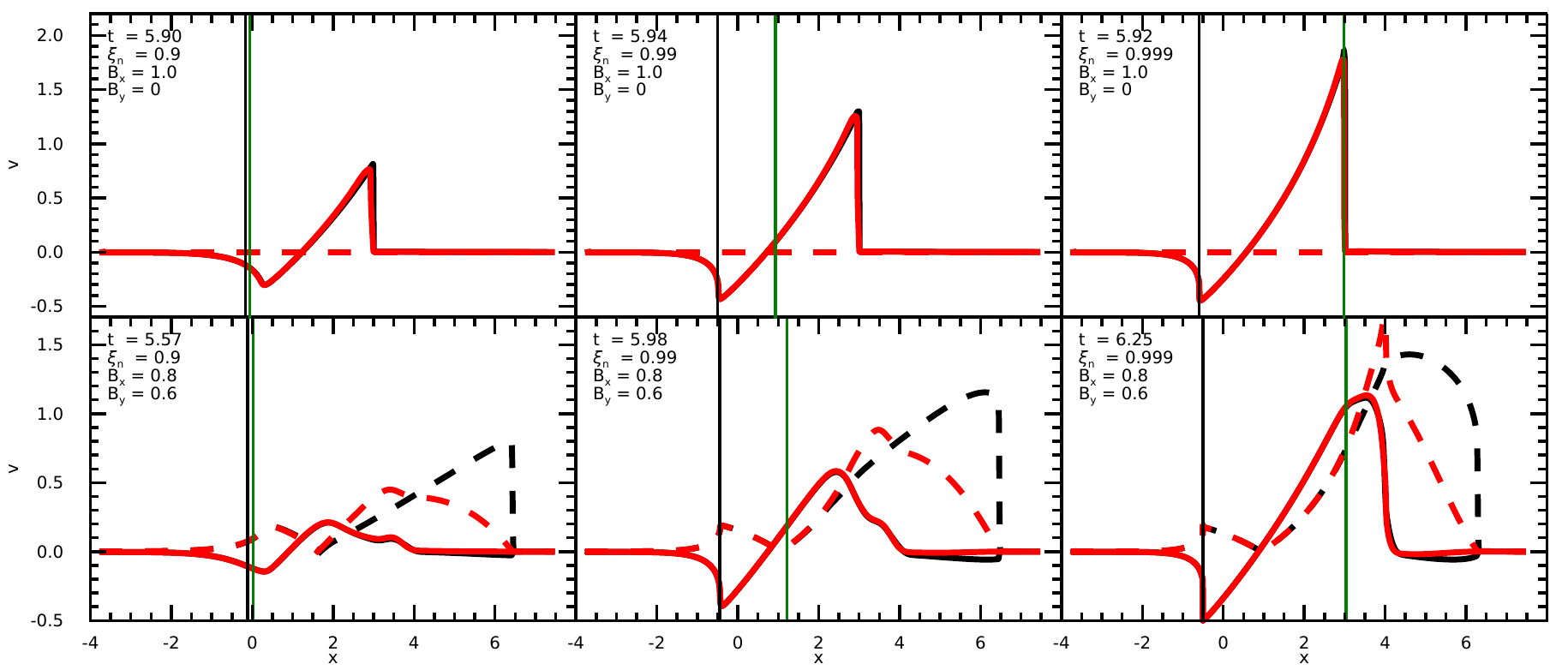}
    \caption{Parallel (solid) and perpendicular (dashed) velocities in the plasma (black) and neutral (red) species for the vertical magnetic field (top row) and inclined field (bottom row) for different neutral fractions ($\xi_n = 0.9,0.99,0.999$ across the columns). The vertical lines indicate the $\beta_v$ (black) and $\beta_{vt}$ (green) points.}
    \label{fig:neutsnap}
\end{figure*}

For the simulations thus far, the collisional coefficient is set as $\alpha _0 = 100$. Changing this value modifies the number of collisions per unit time and hence determines the level of coupling between the neutral and plasma species. 
Taking $\alpha _0 = 0$ would yield the same result as the reference MHD calculation in the plasma since the two species propagate independently. 
At the other extreme, taking $\alpha_0=\infty$ would also yield an MHD result but the wave speeds and the rate at which nonlinearities develop would all depend on the total density ($\rho_p+\rho_n$). 
Between these extremes, the coupling is finite and one would expect two-fluid effects to be most important.
One may also expect separation of wave modes to occur between $\beta _v$ and $\beta _{vt}$.

Simulations have been performed using $B_x=0.8,B_y=0.6$ (corresponding to Section \ref{sec:a1}), using different values of collisional coupling, and a neutral fraction of $\xi_n =0.999$. 
We extend the domain to allow sufficient height above the $\beta _{vt}$ point. The gravity in this extended region is zero. 
This prevents the fast mode wave from accelerating above $x \approx 3.7$. The new atmosphere is shown in Figure \ref{fig:longatmos}. 

Figure \ref{fig:alpsnap} shows the PIP simulations using different collisional coefficients, and the MHD case for reference. The corresponding shock transitions are included in Appendix \ref{app:shocktranscol} for reference. 
The MHD results shows a sharp perpendicular (fast-mode) shock, and a smoothed parallel (slow-mode) component, as discussed in Section \ref{sec:a1MHD}. 
For the weakly collsional case ($\alpha _0 =0.001$), the plasma component of the PIP simulation is very similar to the MHD simulations, in terms of both features and amplitudes. The plasma component is very weakly influenced by the neutral species and therefore they propagate fairly independently and the plasma wave modes separate at the $\beta _v$ point. A fast transition occurs in the perpendicular plasma velocity using the characteristic wave speeds of the isolated plasma. There is no parallel plasma shock transition since this has been smoothed (as seen in the MHD case). In the neutral species, a sonic shock exists in both the parallel and perpendicular components of velocity. The neutral parallel and perpendicular shocks are located at the same height since the neutrals are behaving independently of the plasma, and hence independently of the magnetic field. 

As the collisional coefficient increases ($\alpha _0 = 0.1$) changes start to appear in the plasma component of the PIP simulation, compared to the MHD case; the amplitude of the fast- and slow- mode shocks have increased, and the propagation time has slowed. The increase in shock amplitude and propagation time was described in Section \ref{sec:wavesteep} and is due to the different pressure scale heights of the neutral and plasma species. The same shock transitions exist as for $\alpha _0 = 0.001$ in the neutral and perpendicular plasma velocities. A new slow-mode shock exists in the parallel plasma velocity. The increased coupling of the system has lead to a steepening of the slow-mode component through the increased compressionallity of the system. 

At $\alpha _0 = 1$, we start to see similarities between the plasma and neutral velocities parallel to the instantaneous magnetic field. The parallel plasma velocity corresponds to the slow component and propagates at a comparable speed to the neutral parallel velocity, indicating an increased coupling over the lower $\alpha _0$ cases.
The plasma perpendicular velocity is clearly responding to the neutral shock even for this low value of $\alpha_c$, as evidenced from the change in shape of the shock. This is partially a result of the increased neutral density and temperature post shock increasing the coupling of the plasma to the neutrals through increased collisions.
The perpendicular velocity corresponds to the fast-mode shock and propagates much faster than the parallel velocity. As a result, the neutrals feel less collisions from the plasma as this wave propagates past them, resulting in a much weaker coupling between the species in this wave front compared to the parallel (slow) component. The perpendicular neutral velocity appears unaffected by the plasma.   
The shock transitions for $\alpha _0 = 1$ show a sonic shock in the neutral species and a fast-mode transition in the perpendicular plasma velocity, as seen for $\alpha _0 = 0.1$ (see Appendix \ref{app:shocktranscol}). However, the parallel plasma velocity shows two different transitions depending on the wave speeds chosen: a slow-mode shock ($3 \rightarrow 4$) in the plasma wave speeds, and an intermediate transition ($2 \rightarrow 3$) in the total wave speeds. Physically, we can deduce that the appropriate choice here is the slow-mode shock using the plasma properties since there is no reversal of the magnetic field (which is a requirement for an intermediate transition). This highlights the importance of choosing the appropriate wave speeds in two-fluid plasma.

At $\alpha _0 = 100$, the parallel neutral and plasma velocities are well coupled, and the plasma perpendicular velocity creates a drag on the neutrals driving a neutral perpendicular velocity. This is similar to the results in Section \ref{sec:a1PIP} for a neutral fraction of $\xi_n =0.9$. Here, the plasma perpendicular shock is fairly broad and smoothed (compared to the discontinuous jump in the MHD case). The coupling is sufficient that no shock transitions exist in the parallel neutral or plasma velocity. The slow-mode component in the plasma is smoothed and the coupling transfers this effect to the neutral species, smoothing the parallel sonic shock. A sonic shock exists in the perpendicular neutral velocity, located at an overshoot in the perpendicular neutral velocity, see Figure \ref{fig:alpsnap}. As the coupling increases, the tail end of the fast-mode shock begins to couple with the neutrals, leading to a large frictional heating of the neutral species and a pressurised expansion in the neural species, creating the overshoot in perpendicular neutral velocity.
The perpendicular plasma velocity shows a shock transition when the plasma wave speeds are used but not when the total wave speeds are used. The shock front is decoupled from the neutrals so one would expect that the plasma wave speeds are most appropriate here. This level of coupling presents an interesting case where the slow-mode plasma velocity is strongly coupled to the neutral species, however the fast-mode plasma velocity is only partially coupled to the neutrals. The fast-mode component travels much faster than the slow mode and hence experiences less collisions per unit distance.

Increasing the collisional coefficient further ($\alpha _0 = 1000$) encourages more coupling between the neutral and plasma species. The perpendicular drift velocity between the two species is small and the plasma component is broader than the MHD case. This shows a significant increase in the level of coupling since both the fast- and slow-mode plasma velocities are strongly coupled to the neutral species. Here, no shock transitions exist in the neutral species. In the perpendicular plasma velocity two possible shocks exist depending on the wave speeds taken; an intermediate transition ($2 \rightarrow 3$) for the plasma wave speeds, and a fast-mode shock  ($1 \rightarrow 2$) using the total wave speeds. We can rule out the intermediate shock since there is no reversal of magnetic field and deduce that here the total wave speeds are most appropriate and a fast-mode shock exists. This indicated that the perpendicular plasma velocity is strongly coupled to the neutral species and the system is now behaving like an MHD case. 

Figure \ref{fig:alpsnapwidth} shows the perpendicular shock width for the plasma (calculated as the distance between the locations of the maximum velocity and the maximum gradient of velocity) as a function of the collisional coefficient. The shock width is calculated when the perpendicular velocity has reached $x \approx 6.5$, corresponding to the snapshots in Figure \ref{fig:alpsnap}. For $\alpha _0 < 0.1$ (and the MHD case), the shock is effectively discontinuous. A small numerical shock width exists due to numerical dissipation of approximately $0.02$ pressure scale heights, see the discussion in Appendix \ref{sec:numdif}. For collisional coefficients in this range, the coupling between plasma and neutral species is weak for the fast-mode component and the perpendicular plasma velocity in the PIP simulation is similar to the MHD simulation, see Figure \ref{fig:alpsnap}. Between $0.1 < \alpha _0 \leq 30 $ the perpendicular shock becomes increasingly broad as the fast-mode plasma shock begins to couple with the neutral species. The neutral drag causes a change in the shape of the plasma fast-mode shock.

A sharp increase in shock width occurs for $30 \leq \alpha _0 \leq 100$ with a maximum occurring around $\alpha _0 = 100$. This occurs because the fast-mode shock has not fully separated from the slow-mode component. Performing the $\alpha _0=100$ simulation in an atmosphere that is stratified up to $x=6$ allows the parallel and perpendicular velocities to separate and the new shock width is 1.19 indicating that even with the reduced total collisions in a fully stratified atmosphere, one can expect shock widths of larger than the pressure scale height for finitely coupled systems. 

As the collisional coefficient increases further ($\alpha _0 \geq 1000$) the shock width decreases and the simulation tends towards an MHD simulation with the dynamics determined by the total density.

It should be noted that while these shocks are very broad, they still satisfy the condition for a fast-mode shock to exist, namely a transition from superfast to subfast across the leading edge of the shock. Finite shock widths are a fundamental feature of two-fluid shocks and arise due to the separation of ionised and neutral species at the shock front \citep[e.g.,][]{Hillier2016,Snow2019}. At the leading edge of the shock, the plasma separates from the neutral species and has a superfast to subfast transition. Following this, the collisional coupling with the neutral species creates a drag that shifts the maximum velocity to far behind the leading edge, resulting in a finite width of the shock that can exceed the pressure scale height.

\subsubsection{Shock separation between $\beta _v$ and $\beta _{vt}$}

The different collisional coefficients change the behaviour of the shock propagation, from MHD-like when the collisional coefficient is small ($\alpha _0 < 1$), to a partially-coupled regime ($1 < \alpha _0 \lessapprox 1000$), and to a strongly coupled system ($1000 \lessapprox \alpha _0$) which is again MHD-like but the properties depend on the total density ($\rho_p + \rho_n$). One would expect the mode separation location to also depend on the collisionality of the system. To investigate the height at which the wave modes separate, the location of the maximum velocity is plotted through time for the parallel and perpendicular components for different collisional coefficients, see Figure \ref{fig:shocklocall}. Here the neutral fraction at the base of the domain is $\xi _n = 0.999$, i.e., the atmosphere in Figure \ref{fig:longatmos}. 

In the MHD case, the wave modes separate at the $\beta _v$ point. Similarly, for a weakly coupled system ($\alpha \lessapprox 3$) the coupling is sufficiently weak that the physics is mostly determined by the plasma component such that the separation occurs at approximately the $\beta _v$ point. 

For the highly collisional case ($\alpha = 1000$) the parallel and perpendicular separate above the $\beta _{vt}$ point indicating that the system is strongly coupled and the total densities are determining the physics. This can be considered as MHD-like since there is such strong coupling between the neutral and plasma species. The perpendicular velocity in the neutral species is strongly coupled to the perpendicular plasma velocity. 

Between these limits, the separation happens between the $\beta _v$ and $\beta_{vt}$ points. At $\alpha _0 = 10$, the separation occurs just above the $\beta _v$ point. At $\alpha _0 = 100$ the separation occurs just below the $\beta _{vt}$ point. The finite coupling between the neutral and plasma species causes the two-fluid effects to become important and the point at which wave mode separate is less trivial to determine. 

As with \cite{Pennicott2019}, we find that the mode conversion height is advected with slow-mode shock for a short time after the mode separation, as shown in Figure \ref{fig:shocklocall}. A larger parameter study would be required to analyse the consequences of this since it is likely related to the velocity amplitude.



\subsection{Effect of neutral fraction}

In this section simulations are now performed using three initial neutral fraction of $\xi_n =0.9,0.99,0.999$ at the base of the domain and the collisional coefficient is set to $\alpha _0 = 100$. The domain is extended for all atmosphere with gravity $g=0$ in the extended region, see Figure \ref{fig:longatmos} for the $\xi_n=0.999$ case. Two angles are investigated: vertical field ($B_x=1.0,B_y=0.0$) and inclined field ($B_x=0.8,B_y=0.6$). Snapshots of the data are taken when the slow shock reaches $x=3$ for the vertical field, and when the fast-mode shock reaches $x=6.5$ for the wide inclined field. The parallel and perpendicular velocities for the plasma and neutral species are shown in Figure \ref{fig:neutsnap}.

For the vertical field, increasing the neutral fraction at the base does not significantly change the results. There is increased compression of the of the shock due to the increased coupled pressure scale height since, for this setup, a higher neutral fraction has more total mass.

For the inclined field ($B_x=0.8,B_y=0.6$) the results are more different. Again, there is an increased wave amplitude due to the increased compression. For the parallel (slow) plasma velocity, the main difference in the increased neutral fraction is an increased amplitude. The parallel plasma velocity couples well with the parallel neutral velocity. The perpendicular plasma velocity shows increased coupling with perpendicular neutral velocity for higher neutral fractions, resulting in a broader shock front. This shock broadening is the same result as seen for increased collisional coefficient (see Section \ref{sec:increasedcoll}). The momentum and energy exchange between the plasma and neutral species depend on the density and collisional coefficient. Here the total density increases as the neutral fraction increases so the coupling terms become larger, hence increasing the coupling coefficient and the neutral fraction have similar effects in this set-up.  

\section{Observational consequences}
The results of this paper indicate that the widths of fast-mode shocks in two-fluid stratified atmospheres can regularly exceed the pressure scale height. Relating this study to the solar chromosphere, the temperature is roughly constant, and the pressure scale height is $L_0 \approx 300$ km, and one would expect fast-mode shocks in a finitely-coupled two-fluid medium to be of approximately this width. This is very different to the intuitive image of shocks as a steep discontinuity and provides a potential observable of two-fluid effects in the lower solar atmosphere.  

To relate these results to the solar atmosphere, our results can be dimensionalised using the pressure scale height of the chromosphere ($L_0 \approx 300$ km), and the sound speed ($c_{s0} \approx 8$ kms$^{-1}$), leading to a time scale $t_0 \approx 40$ s. Sampling the Doppler velocity along the direction of stratification ($v_{xp}$) at a fixed height above the $\beta _{vt}$ point, we find that the fast-mode shock should appear as a gradual increase in Doppler velocity over approximately 6 seconds. As such, with high temporal cadence and a narrow spectral line, one would expect this feature to be observationally resolvable. 

In addition, previous studies have shown that another potential observable of two-fluid effects is the electric field felt by the neutrals \citep{Anan2014,Snow2019}. The large drift velocities lead to polarisation of the neutrals as they flow past the magnetic field, and hence have an associated electric field term of the form $\textbf{E}_n=(\textbf{v}_n-\textbf{v}_p) \times \textbf{B}$. The large drift velocities (up to $80 \%$ of the plasma sound speed) present over the large shock width in the fast-mode shock provide an excellent scenario for observing the electric field felt by the neutrals.

\section{Conclusions}
In this paper, we have investigated the propagation and mode conversion of two-fluid shocks in an isothermal, stratified atmosphere. The collisionallity of the system is found to have a strong effect on determining the behaviour of the system. Two angles of magnetic field are investigated: vertical ($B_x=1,B_y=0$), and inclined ($B_x=0.8,B_y=0.6$). 

There are several quirks of two-fluid simulations that are important in this study. Firstly, the neutral and ionised species have different pressure scale heights, meaning that a coupled two-fluid wave propagates slower and steepens more than an isolated MHD wave. Secondly, the characteristic wave speeds of a two-fluid plasma are multiply defined depending on the level of coupling, see Section \ref{sec:wavespeeds}. Finally, the point at which mode conversion occurs is multiply defined: $\beta _v$ ($c_s=v_A$) using the MHD (or plasma) properties, and $\beta _{vt}$ ($c_{st}=v_{At}$) using the properties of the coupled system (plasma + neutrals).

For a vertical magnetic field ($B_x=1,B_y=0$) the initial wave steepens into a slow-mode shock. When the shock passes the $\beta _v$ point in the MHD simulation, there is no significant changes in the shock since the propagation is restricted to along the magnetic field direction. In the PIP case (for $\xi_n=0.9$), there is a separation of the neutral and plasma species that appears at the leading edge of the shock above the $\beta _v$ point. 
The plasma sound speed is larger than the neutral sound speed by a factor of $\sqrt{2}$ hence the plasma shock propagates ahead of the neutral shock.
The frictional coupling between the two species forces the neutral species to propagate only slightly behind the plasma, and there is an associated increased temperature of the neutral species (Figure \ref{fig:PIPstraightheating}).
Increasing the neutral fraction increases the amplitude of the shock due to the increased compressionality of the system. The propagation speed of the shock remains the same since the sound speed does not change by increasing the neutral fraction.

When the magnetic field is highly inclined ($B_x=0.8,B_y=0.6$) the initially coupled wave separates into fast- and slow-mode components above the $\beta _v$ point. In the MHD case, this results in a smoothing of the parallel (slow) component of the velocity (see Figure \ref{fig:straighttimeseries}), as analysed by \cite{Pennicott2019}. For the two-fluid PIP case (for $\xi_n=0.9, \alpha_0 = 100$), the smoothing of the parallel (slow) shock component results in a strong coupling between the neutral and plasma species due to the smoothing and does not possess a shock transition in either the neutral or plasma species. 
The perpendicular (fast) component of velocity exhibits a reasonably strong drift velocity near the shock front. The neutral species does not possess an MHD fast wave speed (since it is hydrodynamic and only supports sonic shocks). The perpendicular component of the neutral species arises from the coupling with plasma fast shock. As such, there is a lag in the peak velocities of the plasma and neutral perpendicular velocities (Figure \ref{fig:bothtimeseriesa1}) and strong increase in neutral temperature due to the frictional heating (Figure \ref{fig:PIPheatinga1}). 

Increasing the neutral fraction and collisional coefficient drastically changes the physics of the system, see Section \ref{sec:colleffect}. For $\xi_n=0.999$, the collisional coefficient results in a MHD-like decoupled system ($\alpha _0 \leq 0.1$), a finitely-coupled system ($0.1 \leq \alpha _0 \leq 100$), and finally a MHD-like coupled system ($100 \leq \alpha _0$). In the extremes of low $\alpha _0 \leq 0.1$, the plasma and neutral species propagate fairly independently, with the plasma fast-mode virtually unaffected by the neutrals, and the plasma slow-mode only weakly affected. The fast-mode propagates much more quickly that the slow-mode and hence experiences fewer collisions per unit distance. As $\alpha _0$ increases, the parallel plasma and neutral species become more coupled. The perpendicular plasma shock becomes increasingly broad as it experiences more collisions with the plasma and results in a shock width of larger than the pressure scale height. For $\alpha _0 = 1000$, both the parallel and perpendicular plasma velocity modes are well coupled to the neutral species and the system becomes MHD-like.

The location at which mode-conversion occurs also depends on the collisionallity of the system. In the MHD case and PIP $\alpha_0 \leq 1$, the fast- and slow- modes separate at the $\beta _v$ point since the dominant separation is determined by the isolated plasma. For a $\alpha_0=1000$ the modes separate above the $\beta _{vt}$ point since the system is strongly coupled and MHD-like. between these two limits, the system is finitely coupled and the mode conversion occurs between the $\beta_v$ and $\beta_{vt}$ points.

Relating our results to the solar chromosphere, one would expect fast-mode shocks in finitely-coupled, two-fluid mediums to have a width comparable to the pressure scale height ($\approx 300$ km in the chromosphere). Such broad shocks could be a indirect way of observing two-fluid effects in the lower solar atmosphere. 

In summary, two-fluid shocks in an isothermal stratified atmosphere feature several peculiarities compared to MHD shocks. The mode-conversion point, shock width and the shock transitions present all depend on the collisionality of the system. The results presented in this paper aid in the understanding of shocks in the lower solar atmosphere.

\section*{Acknowledgements}
BS and AH are supported by STFC research grant ST/R000891/1. AH acknowledges the support of STFC Ernest Rutherford Fellowship grant number ST/L00397X/2. 
The authors wish to acknowledge scientific discussions with the Waves in the Lower Solar Atmosphere (WaLSA; \url{www.WaLSA.team}) team,  which  is  supported  by  the  Research  Council  of  Norway  (project  number  262622).


\bibliographystyle{aasjournal} 
\bibliography{PIP_smoothing} 
   
\appendix

\section{Numerical implementation of velocity driver} \label{driver}
The PIP code evolves conserved quantities. As such, the numerical implementation of the driver described by Equations \ref{eqn:driver1}-\ref{eqn:driver3} in normalised form is:
\begin{gather}
    c_{s0} = \sqrt{\gamma P_0 / \rho_0}, \\
    v_{pert} = A \sin{(2\pi t/\tau)}, \\
    v_{pert}' = \frac{2\pi A}{\tau} \cos{(2\pi t/\tau)},  \\
    P_{pert}' = \frac{P_0 \gamma}{c_{s0}} v_{pert}', \\
    \rho_{pert} = \frac{\rho_0}{c_{s0}} v_{pert}, \\
    \rho_{pert}' = \frac{\rho_0}{c_{s0}} v_{pert}', \\
    \rho = \rho + \rho_{pert}' dt, \\
    \rho v_x = \rho v_x + (\rho_{pert}' v_{pert} + (\rho_{pert} + \rho_0)v_{pert}') dt, \\
    e = e + \left(\frac{1}{2} v_{pert}^2 \rho_{pert}' +(\rho_{pert}+\rho_0) v_{pert} v_{pert}' + \frac{P_{pert}'}{(\gamma -1)} \right)dt,
\end{gather}
where $dt$ is the simulation time step, and $\rho, \rho v_x, e$ are the evolved simulation density, $x$ momentum and energy on the driven boundary.

\section{Numerical diffusion} \label{sec:numdif}

\begin{figure}
    \centering
    \includegraphics[width=0.495\textwidth,clip=true,trim=1.2cm 8.0cm 1.4cm 8.4cm]{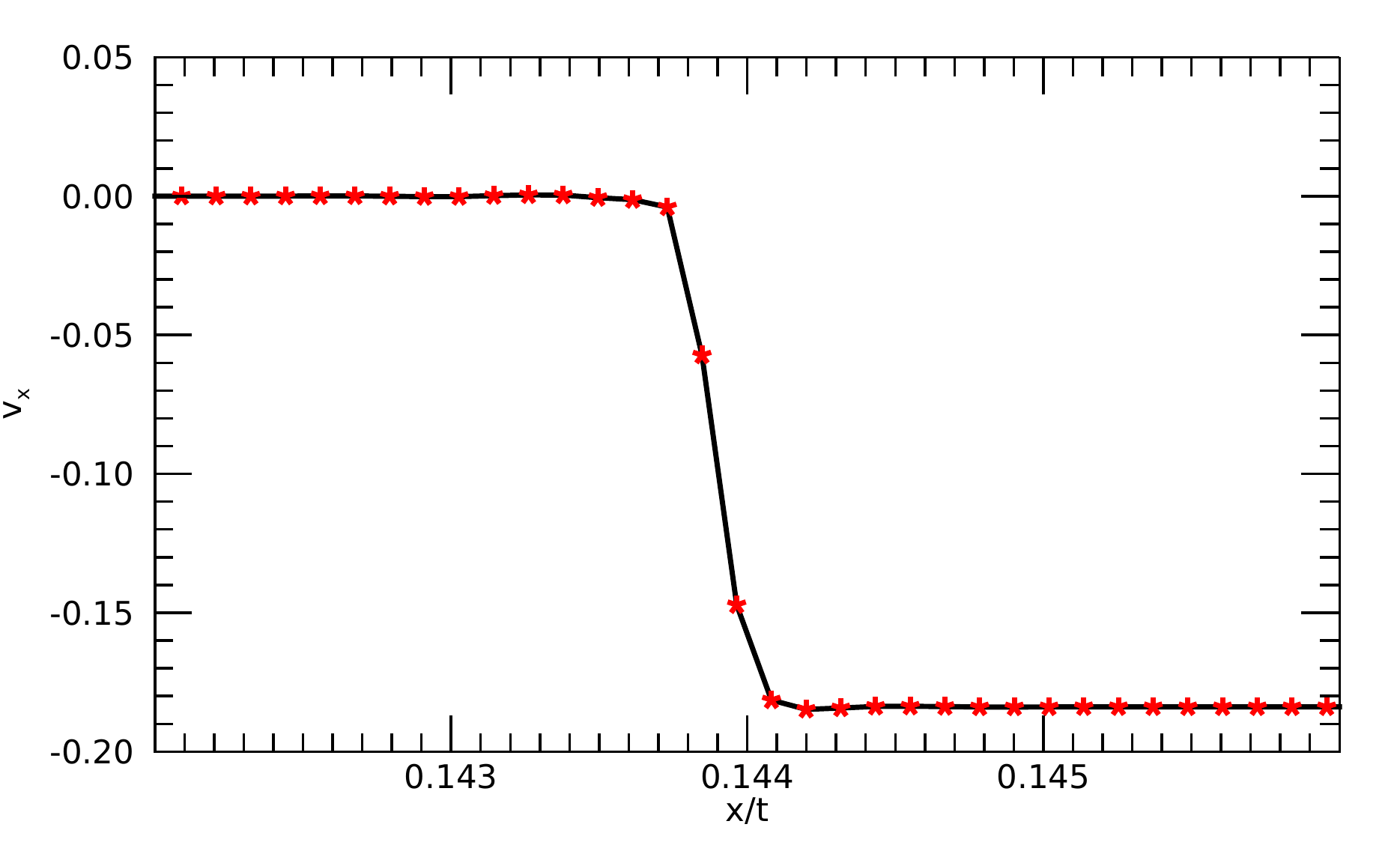} \\
    \includegraphics[width=0.495\textwidth,clip=true,trim=1.2cm 8.5cm 1.4cm 8.4cm]{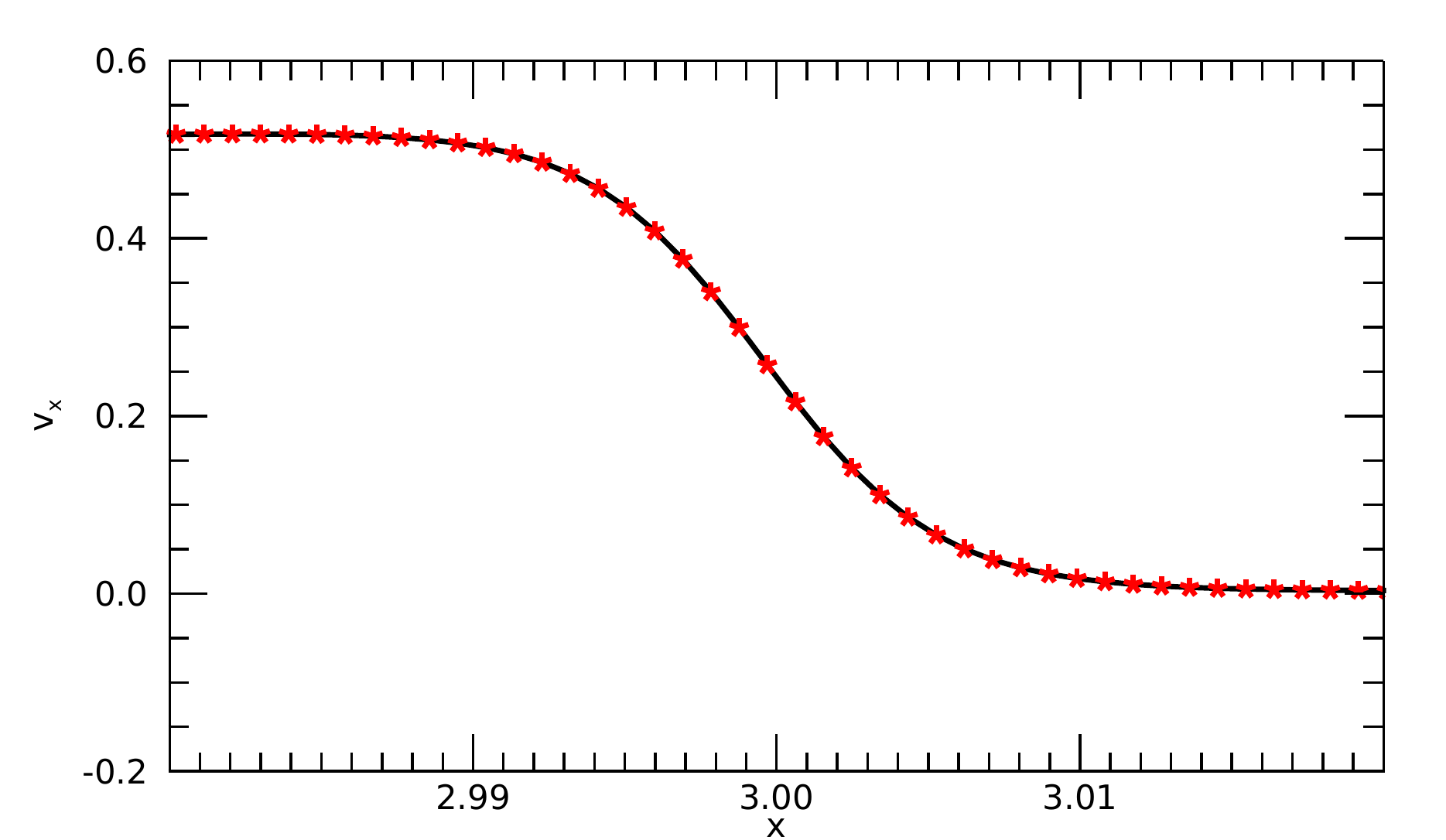}
    \caption{Numerical diffusion tests for a slow-mode shock in non-stratified (top) and stratified (bottom) atmospheres. The top panel corresponds to an MHD slow shock used in \cite{Hillier2016} and the bottom panel corresponds to the MHD simulation in Figure \ref{fig:straighttimeseries} at time $t=5.68$. The numerical shock width in the stratified atmosphere is less then $0.02$ pressure scale heights and considered negligible.}
    \label{fig:difftest}
\end{figure}

We use a first order HLLD numerical scheme, as described by \cite{Miyoshi2005}. The (P\underline{I}P) solver have been used with this scheme previously to model MHD shocks in a non-stratified atmosphere \citep{Hillier2016,Snow2019} and the numerical diffusion of the scheme was found to be small, with the shock existing across a few grid cells, as shown in Figure \ref{fig:difftest}. In the stratified atmosphere used in this paper, the shock is found to be more diffuse since the densities either side of the shock are constantly changing and hence a true steady-state solution does not exist. This is demonstrated in Figure \ref{fig:difftest} which shows the MHD slow-mode shock for a vertical magnetic field and corresponds to the MHD simulation in Figure \ref{fig:straighttimeseries} at time $t=5.68$). In the stratified case, the physical shock width is approximately 0.02 pressure scale heights and hence we consider the numerical diffusion a negligible effect.

\section{Shock transitions for the collisional tests} \label{app:shocktranscol}

\begin{figure}
    \centering
    \includegraphics[width=0.5\textwidth,clip=true,trim=5.4cm 2.2cm 6.6cm 1.4cm]{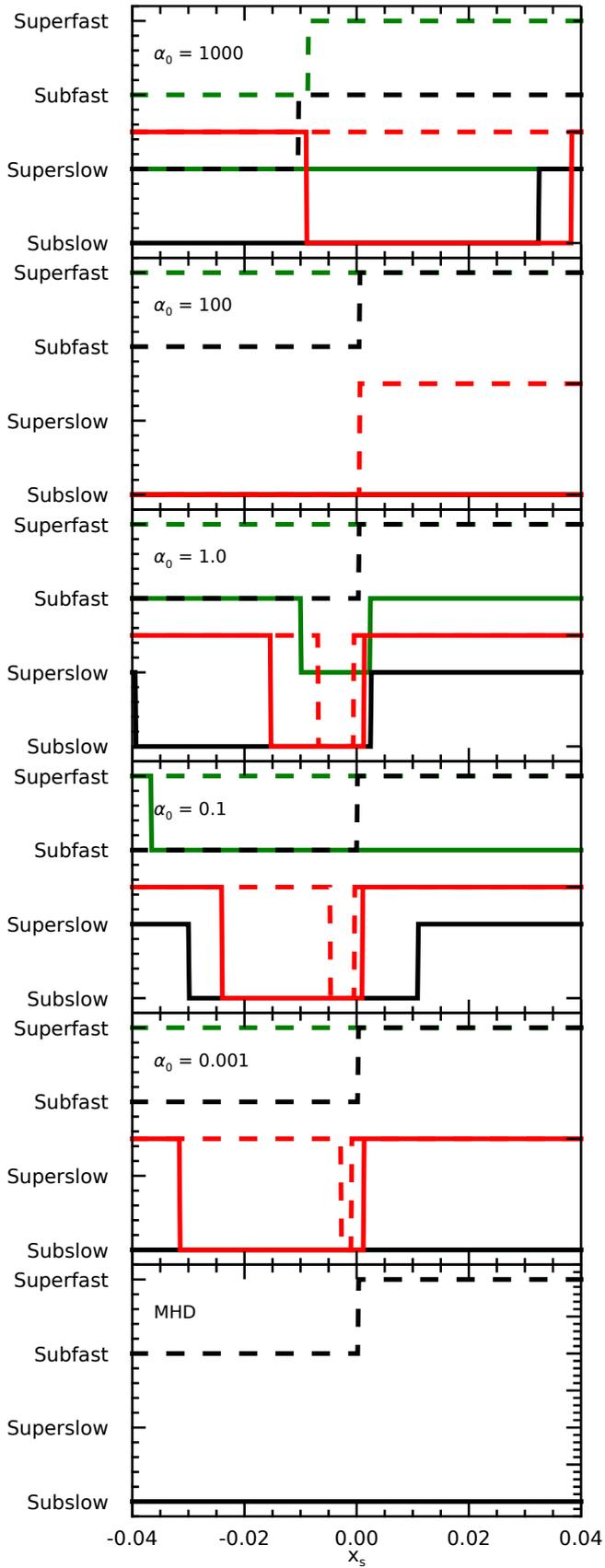}
    \caption{Shock transitions for different collisional coefficients. Parallel (solid) and perpendicular (dashed) lines are shown for the plasma (black and green) and neutral (red) species. The black lines show the transitions relative to the isolated plasma, whereas the green lines show the transitions relative to the fully coupled system.}
    \label{fig:shocktranscol}
\end{figure}

Figure \ref{fig:shocktranscol} shows the shock transitions for the different collisional fraction in Section \ref{sec:colleffect}. 

\end{document}